\documentclass[runningheads]{llncs}
\usepackage{graphicx}
\usepackage{booktabs}
\usepackage{tabularx}
\newcommand\setrow[1]{\gdef\rowmac{#1}#1\ignorespaces}
\usepackage[table]{xcolor}
\usepackage{threeparttable}
\usepackage{multirow, makecell, caption}

\usepackage[algo2e,boxed,algoruled]{algorithm2e}
\usepackage{amsmath}
\usepackage{amssymb}
\usepackage{xcolor}
\usepackage{subfig}
\usepackage{authblk}
\begin{document}
\title{A training-free recursive multiresolution framework for diffeomorphic deformable image registration}
%
%
\def\correspondingauthor{*Corresponding author. E-mail: sheikhja@ualberta.ca }
\def\equalcont{**These authors contributed equally to this work.}
\def\contributauthor{Contributing authors: mnoga@ualberta.ca,punithak@ualberta.ca,nray1@ualberta.ca}
\author{Ameneh Sheikhjafari\inst{1,3*}\and Michelle Noga\inst{2,3} \and Kumaradevan Punithakumar\inst{2,3**} \and Nilanjan Ray\inst{1**}}
\institute{ Department of Computing Science, University of Alberta, Canada \and Radiology and Diagnostic Imaging, University of Alberta, Canada \and Servier Virtual Cardiac Centre,  Mazankowski Alberta Heart Institute, Canada\\
\correspondingauthor{}\\
\contributauthor{}\\
\equalcont}

%
\maketitle              
\begin{abstract}
Diffeomorphic deformable image registration is one of the crucial tasks in medical image analysis, which aims to find a unique transformation while preserving the topology and invertibility of the transformation. Deep convolutional neural networks (CNNs) have yielded well-suited approaches for image registration by learning the transformation priors from a large dataset. The improvement in the performance of these methods is related to their ability to learn information from several sample medical images that are difficult to obtain and bias the framework to the specific domain of data. In this paper, we propose a novel diffeomorphic training-free approach; this is built upon the principle of an ordinary differential equation.
 
Our formulation yields an Euler integration type recursive scheme to estimate the changes of spatial transformations between the fixed and the moving image pyramids at different resolutions. 
The proposed architecture is simple in design. The moving image is warped successively at each resolution and finally aligned to the fixed image; this procedure is recursive in a way that at each resolution, a fully convolutional network (FCN) models a progressive change of deformation for the current warped image. The entire system is end-to-end and optimized for each pair of images from scratch. In comparison to learning-based methods, the proposed method neither requires a dedicated training set nor suffers from any training bias. We evaluate our method on three cardiac image datasets. The evaluation results demonstrate that the proposed method achieves state-of-the-art registration accuracy while maintaining desirable diffeomorphic properties.

\end{abstract}
\section{Introduction}
Deformable image registration is an important part of a variety of medical image processing applications. For instance, deformable registration is applied for the quantification of regional function measures and tracking heart motion~\cite{khalil2018overview,sheikhjafari2015robust} in cardiology. 
The aim of the deformable image registration is to establish a point-wise correspondence between two images to align them by estimating an appropriate non-linear transformation. 
 
There is a strong need for the development of accurate and reliable nonrigid registration algorithms for many medical imaging applications. 
Many deformable registration algorithms have been proposed by various research groups and existing algorithms could be broadly categorized into non-learning-based (classical) methods and learning-based methods. 

Commonly, non-learning-based registration techniques solve regularized optimization problems using an image similarity/dissimilarity metric and regularization. These methods include nonlinear elastic-based methods~\cite{burger2013hyperelastic}, diffusion-based methods~\cite{lorenzi2013lcc}, B-splines \cite{marstal2016simpleelastix,sheikhjafari20153d}, demons algorithm~\cite{mansi2011ilogdemons}, diffeomorphic transformation-based methods \cite{avants2008symmetric,lorenzi2013lcc,punithakumar2017gpu}. Adapting the conventional methods to specific problems is challenging, as they require a precise model parameterization that can be too flexible or too restrictive. Besides, occasionally regularization terms need to be manually adjusted for each application to obtain accurate results. However, recent studies show that the regularization prior can be modeled by neural networks, independent of learning, for deformable image registration \cite{laves2019deformable,sheikhjafari2018unsupervised} to remove the need for manual adjustment of parameters.

In the past few years, learning-based approaches, notably deep CNNs, have been successfully applied in different computer vision problems such as color enhancement, color compensation, and image registration \cite{de2017end,xu2020color,xu2020multi,xu2021novel,yan2020multistep}. 
In supervised-learning based image registration methods, CNN is trained using examples of medical images along with their ground truth transformations to predict the transformations directly on test images \cite{cao2018deformable,rohe2017svf,yang2017quicksilver}. Even though the accuracy of these approaches is considerable, their performance is bounded by the quality of the synthetic ground truth deformation field or the annotated map~\cite{sang2020imposing}. Additionally, the actual ground truth of a desired neural network output is not often available. 

Unsupervised image registration methods have received a lot of attention because they bypass the need for annotated data. In addition, the quality of the output in unsupervised registration methods does not depend on the quality of labels. However, they are often trained on a set of images which could introduce bias on specific data. These methods utilize a CNN and a differentiable similarity function to learn the dense spatial mapping between input image pairs \cite{balakrishnan2018unsupervised,de2017end}. Also, it must be noted that most of the existing CNN-based methods often do not consider desirable diffeomorphic properties including invertibility and inverse consistency of the transformation~\cite{mok2020fast}.

In this paper, we propose a novel learning-free diffeomorphic recursive framework using an FCN capable of estimating plausible non-linear transformations between images. The main elements of this work are:
\begin{enumerate}
    \item In the proposed method, an FCN models \textit{the changes} in the deformation over multiple resolutions as opposed to the deformation itself that most existing methods do. The final deformation is estimated by a solution to an ordinary differential equation (ODE). Thus the resulting algorithm is recursive in nature.
    \item Following this recursion, the moving image is warped successively, enabling the final prediction, which might consist of large displacements, to be decomposed into smaller displacements. 
    \item Within the recursive algorithmic framework, we introduce diffeomorphism, which guarantees the inverse consistency of deformations.
    \item The proposed method is learning-free, i.e., it optimizes the parameters of the FCN from scratch for every new pair of images and eliminates the need for a dedicated training set. 
    \item The parameters of the FCN are shared across all resolutions and the neural net loss function values at all resolutions are optimized simultaneously.
\end{enumerate}

\begin{algorithm2e}[!t]
\caption{ODE-based Multi-resolution Diffeomorphism}
\label{alg:net2}
\KwIn{Pixel coordinates $\{X_t\}_{t=1}^{K}$ for all resolutions}
\KwOut{Deformation fields $\{D_t\}_{t=1}^{K}$ for all resolutions}
\DontPrintSemicolon
-Initialize a spatial transformation\;
 \qquad $V_{K} = f_{\theta}(X_{K},0)$\;
 \qquad $D_{K} = EXP(V_K)$\;
\For{$t\leftarrow K-1$ \KwTo $0$}{
      $D^{up}_{t+1} = UpSample(D_{t+1})$\;
      $V_t = f_{\theta}(X_t,D^{up}_{t+1})$\;
      $V_t \leftarrow LowPass(V_t)$\;
      $D_{t} = D^{up}_{t+1} \circ  EXP(V_t)$\;
    }
\end{algorithm2e}%

\section{Methodology} 
Let $I_m\in \Omega$, $I_f \in \Omega$ denote the moving image and the fixed image respectively which are both defined over d-dimensional space $\Omega$. $D$ is the deformation field which maps $D: \Omega \xrightarrow{} \Omega$. Commonly, deformable registration methods construct a deformation prediction function $F$ which takes moving and fixed images as inputs and predicts a dense deformation field that aligns $I_m$ to $I_f$. 
In contrast, in the proposed approach, we construct $F$ as a function of coordinate grid (not images) to predict the the change in the deformation field from one resolution to the next resolution. 
We cascade this procedure by recursively performing registration on the multi-resolution levels. Following this recursion, the change of deformation field is predicted at each resolution, enabling the final deformation field probably with large displacement to be decomposed into cascade with small displacement.

In multi-resolution pyramids, a Gaussian pyramid of images is constructed where the original image lies at the bottom level and subsequently higher levels have a down-scaled Gaussian blurred version of the image. Multi-resolution pyramids often serve to accelerate the optimization and yield better accuracy. Using a multi-resolution recipe, two image pyramids are built: $I_F^t$ and $I_M^t$ for $t=K,\dots,1$, where K is the maximum level in the pyramid. Here, $I_F^1 = I_F$ and $I_M^1 = I_M$ are the original fixed and moving images, respectively.

Considering resolution $t$ as the continuous variable, the change in deformation over resolution $t$ is conceived by a FCN $f_\theta$:
\begin{equation}
    \frac{d}{dt}D_t=f_\theta(X_t,D_t),
\end{equation}
where $X_t$, a multi-dimensional array, is the pixel coordinate grid at resolution $t$ and $D_t$ is the deformation field, another multi-dimensional array at resolution $t$. Thus, we model the change in deformation as a function of both the coordinate grid and the deformation. Discretization of the resolution (i,e., pyramid levels) leads to a solution by a Euler integration with an initial value $D_K$:
\begin{equation}
    D_{t}=D_{t+1}+f_\theta(X_{t+1},D_{t+1}), ~~\text{for}~ t=K-1,K-2,...,1,
\end{equation}
where $t=K$ is the coarsest and $t=1$ is the finest/original resolution. $D_{t+1}$ is the upsampled deformation field from a lower level $D_{t}$ and $f_\theta(D_{t+1})$ is the changes of the deformation field at resolution $t+1$ from resolution ${t}$.

Additionally, to have the same canonical range of pixel coordinates at every resolution, in our implementation, we use the range $[ -1, 1] \times [ -1, 1]$ for pixel coordinates. With this view, a multi-resolution pyramid adds more samples in the space $[ -1, 1] \times [ -1, 1]$ as we go from lower to higher resolutions. 

A reasonable deformation field should prevent folding and be continuously varying (diffeomorphism). By taking the composite form in the Euler integration instead of the additive form, we introduce diffeomprhism to the framework. Further, if we treat the output of the FCN as the velocity field, then exponential of the velocity field will yield the multi-resolution diffeomorphic deformation:
\begin{equation}
    D_{t}=D_{t+1} \circ EXP(f_\theta(X_{t+1},D_{t+1})), ~~\text{for}~ t=K-1,K-2,...,1.
\label{eq:euler}
\end{equation}

The computational scheme for (\ref{eq:euler}) appears in Algorithm \ref{alg:net2}, where we introduced two more functionalities: upsampling and smoothing. Upsampling interpolates a deformation from a lower to the next upper level of resolution and smoothing of the velocity field acts as an implicit regularizer for image registration. 

\begin{figure*}[!t]
\centering
    \scalebox{0.26}{
    \includegraphics{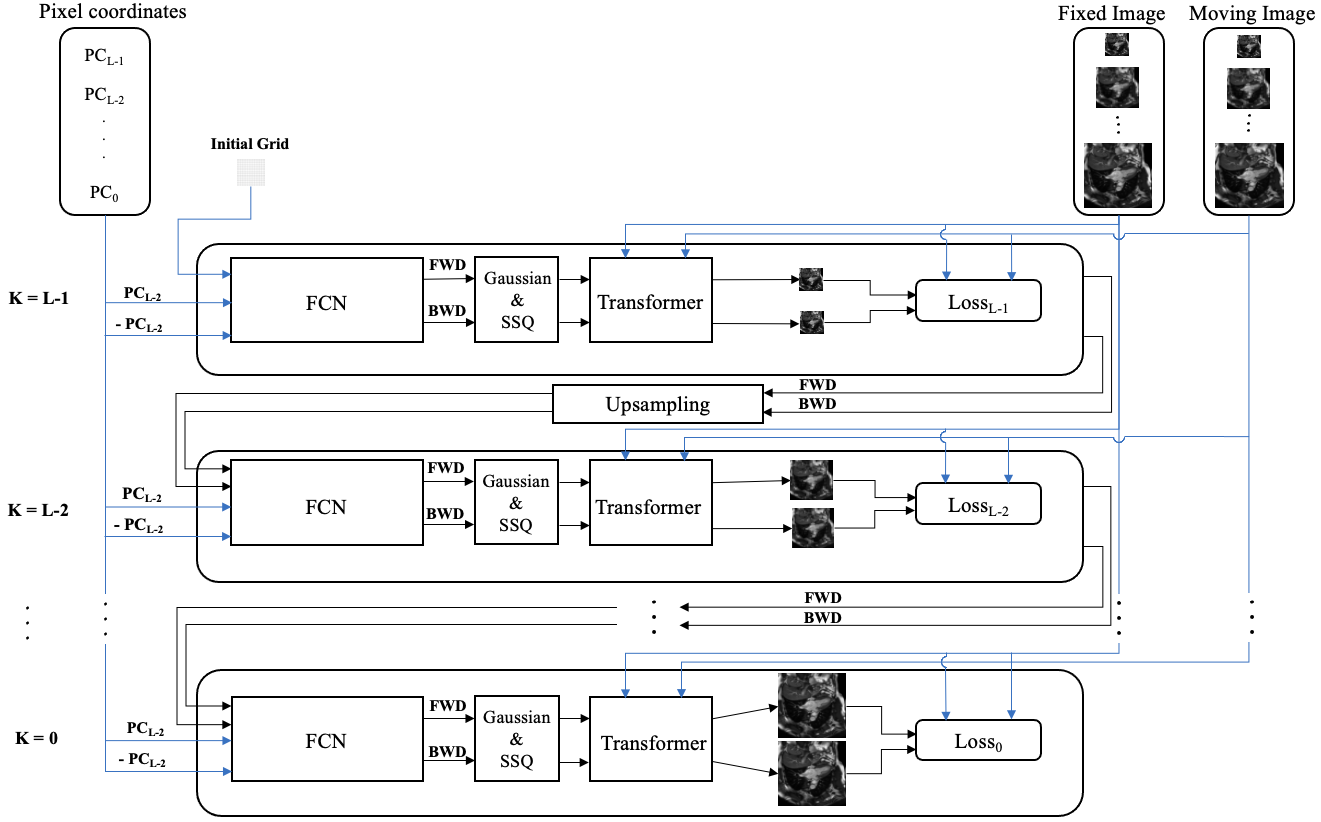}}
	\caption{Schematic of the proposed recursive algorithm.
	$PC_{K-1}, PC_{K-2}, ... , PC_0$ are pixel coordinates correspond to the image pyramids. $PC$ and $-PC$ are passed as the inputs to the network for forward and backward directions. An initial grid with zero displacement is used as a correction for the resolution $t=K-1$. The outputs of the FCN, forward deformation field $FWD$ and backward deformation field $BWD$ will be smooth by a Gaussian filter and then passed to the scaling and squaring method $SSQ$. The final $FWD$ and $BWD$ generated at each resolution will be passed to an upsampler and then the upsampled results will be used as a correction for the next resolution. For each pair of images we repeat the whole process (e.g. $800$ iterations) till we find the optimum value of $\theta$}
	\label{fig:net}
\end{figure*}

In this work, we use geometric optimization that uses the local canonical coordinates \cite{vercauteren2007non}. Based on this method, to estimate the current deformation, a compositional update rule is used between the previous estimate and the exponential map $EXP$ of the displacement field. The exponential map is efficiently calculated by using the scaling and squaring method \cite{higham2009scaling} and the composition of displacement fields. The exponentiation of the displacement field ensures the diffeomorphism of the mapping. Therefore, our recursive diffeomorphic image registration is obtained with Algorithm~\ref{alg:net2}. Also the schematic of the proposed recursive algorithm is shown in \ref{fig:net}. 
To optimize the FCN parameters $\theta$, we use two Gaussian pyramids, one each for the moving $I_M^t$ and the fixed $I_F^t$ images, where $t=1,...,K$. Then, the FCN's parameters will be optimized based on the difference between warped moving image and fixed image at the same resolution. In our approach, the neural network will be optimized from the scratch for each pair of images being registered, and therefore, the trainable parameter values of the network will at convergence will be different. Our approach is a classical learning-free setting which would not require a dedicated training set with annotations of any kind from experts.

The parameters $\theta$ of the FCN are found by solving the following optimization using gradient descent (we use PyTorch's autograd):
\begin{equation}
    \label{eqmin1}
    \begin{split}
    \min_{\theta} \sum_{t=1}^{K}\sum_{x\in \Omega} & L(I_{F}^t(x),I_{M}^t(D_{t}(x))) + \gamma\left\lVert D_{t}(x)-x\right\rVert^{2},
    \end{split}
\end{equation}
where $\Omega$ denotes the fixed and moving images domain, $I_{F}^t(x)$ is the fixed image pixel value at pixel location $x$ and resolution level $t.$ $I_{M}^t(D_t(x))$ in (\ref{eqmin1}) refers to the pixel value on the moving image $I_{M}^t$ at location $x$ after being displaced by deformation field $D_t(x).$ $L(.,.)$ is a differentiable loss function. In order to use the same range of displacement for all levels of resolution, we normalized the deformation field and the index at all resolutions in a canonical range $[-1,1]$. 
\begin{figure*}[!t]
	  \setlength{\belowcaptionskip}{-8pt}
		\begin{minipage}[b]{0.190\linewidth}
		\centering

		\centerline{\includegraphics[width=\textwidth]{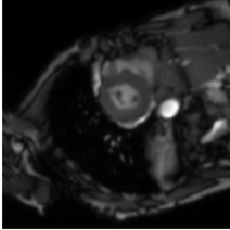}}
		\centerline{ES}\medskip
	\end{minipage}
	\hfill
	\begin{minipage}[b]{0.190\linewidth}
		\centering

		\centerline{\includegraphics[width=\textwidth]{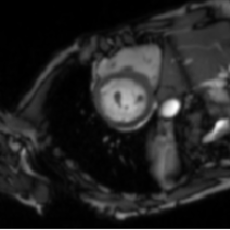}}
		\centerline{ED}\medskip 
	\end{minipage}
	\hfill
	\begin{minipage}[b]{0.190\linewidth}
		\centering

		\centerline{\includegraphics[width=\textwidth]{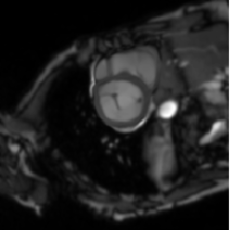}}
		\centerline{Registered ES}\medskip
	\end{minipage}
	\hfill
	\begin{minipage}[b]{0.190\linewidth}
		\centering

		\centerline{\includegraphics[width=\textwidth]{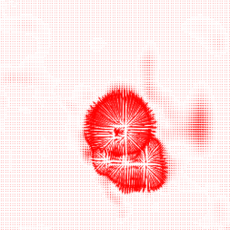}}
		\centerline{Flow}\medskip
	\end{minipage}
	\hfill
	\begin{minipage}[b]{0.190\linewidth}
		\centering

		\centerline{\includegraphics[width=\textwidth]{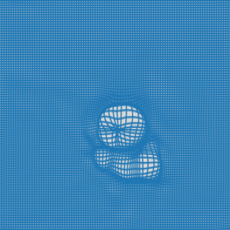}}
		\centerline{$D^F$}\medskip
	\end{minipage}
	
		\begin{minipage}[b]{0.190\linewidth}
		\centering

		\centerline{\includegraphics[width=\textwidth]{J.png}}
		\centerline{ED}\medskip
	\end{minipage}
	\hfill
	\begin{minipage}[b]{0.190\linewidth}
		\centering

		\centerline{\includegraphics[width=\textwidth]{I.png}}
		\centerline{ES}\medskip 
	\end{minipage}
	\hfill
	\begin{minipage}[b]{0.190\linewidth}
		\centering

		\centerline{\includegraphics[width=\textwidth]{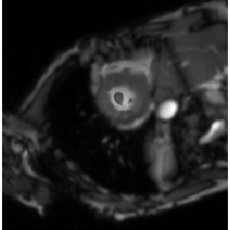}}
		\centerline{Registered ED}\medskip
	\end{minipage}
	\hfill
	\begin{minipage}[b]{0.190\linewidth}
		\centering

		\centerline{\includegraphics[width=\textwidth]{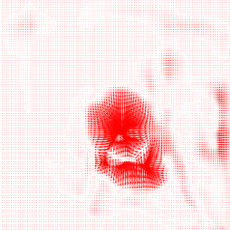}}
		\centerline{Flow}\medskip
	\end{minipage}
	\hfill
	\begin{minipage}[b]{0.190\linewidth}
		\centering

		\centerline{\includegraphics[width=\textwidth]{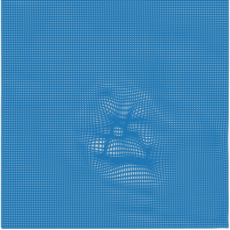}}
		\centerline{$D^B$}\medskip
	\end{minipage}
	\caption{A sample result of forward and backward registrations with velocity fields and grid deformation fields from ACDC dataset.}
	\label{fig:forwardandback}
\end{figure*}

In most of the learning-based deformable image registration approaches, the inverse mapping is often ignored~\cite{mok2020fast}. However, the accuracy of the nonrigid registration could be improved by computing the mappings from moving to fixed image and vice versa as well as exploiting the invertible property of diffeomorphic transformations.

In the proposed registration framework, we employ a FCN with two dense velocity fields to generate two deformation fields, namely, forward $D^F$ and backward $D^B$ deformation fields, that register images $I_M$ to $I_F$ and $I_F$ to $I_M$, respectively, using Algorithm~\ref{alg:net2}. We then enforce the consistency constraint that these two transformations are inverse mappings of each other by adding a regularization term to the loss function. Our FCN architecture consists of 4 convolution layers, each contains twenty-four $5 \times 5$ filters with a stride of $1$, followed by a rectified linear unit (ReLU) activation, except for the last layer.

Note that the output of the Algorithm~\ref{alg:net2} is $D_t=[D_t^F, D_t^B]$ in which $D_t^F$ and $D_t^B$ are forward and  backward deformation fields, respectively, see Fig.~\ref{fig:forwardandback} as an example. With this view the following forward-backward optimization is used:
\begin{equation}
    \label{eqmin2}
    \begin{split}
    \min_{\theta} &\sum_{t=1}^{K}\sum_{x\in \Omega}[
    L_1(I_{F}^t(x),I_{M}^t(D_{t}^F(x))) +
   L_1(I_{M}^t(x),I_{F}^t(D_{t}^B(x))) + \\
    &\alpha L_2(D_t^B(D_t^F(x)),x) +
     \alpha L_2(D_t^F(D_t^B(x)),x) +\\
    &\gamma\left\lVert D_t^F(x)-x\right\rVert^{2} + \gamma\left\lVert D_t^B(x)-x\right\rVert^{2}]
    \end{split}
\end{equation}%
We use differentiable mutual information (MI) \cite{belghazi2018mine} and structural similarity index metric (SSIM) \cite{belghazi2018mine} as similarity metric $L_1 = MI + SSIM$ and mean squared error (MSE) as penalizes term to enforce the inverse consistency $L_2 = MSE$. 

\begin{table}[!th]
\centering
\caption{Quantitative evaluation of the results for cardiac MRI registration on the ACDC, SCD and LA datasets. The evaluation was performed in terms of Dice (mean$\pm$ standard deviation) and HD (mean) and the average number of pixels with non-positive Jacobian Determinant (lower is better). Values in bold indicate the best performance.}
  \begin{minipage}{1\linewidth}
  \centering
  \begin{threeparttable}
    \begin{tabularx}{1\textwidth}{ l >{\centering\arraybackslash}X r c l >{\centering\arraybackslash}X c c}
    \toprule
    Methods&&ACDC&\multicolumn{2}{c}{}&&SCD\\
    \cmidrule(lr){2-4} \cmidrule(lr){5-8}
         & Dice & HD & $\det(J_{\phi})<0$  & &Dice & HD & $\det(J_{\phi})<0$\\
    \midrule
    \small{Unregistered} & 0.65 $\pm$ 0.20 & 17.76  & N/A && 0.62  $\pm$ 0.15 & N/A &N/A  \\
    \small{RTT(L2L2)} & 0.76 $\pm$ 0.18 & 7.19 & 0.10&& 0.70 $\pm$ 0.12 & 5.36 & 0.30 \\
    \small{RTT(L2L1)} & 0.78 $\pm$ 0.18 & 6.64 & 0.25 && 0.70 $\pm$ 0.12 & 5.36 & 0.30 \\
    \small{Elastix}  &  0.79 $\pm$ 0.18 &  11.26 & 0.32 & &  0.79 $\pm$ 0.08 &  N/A & 0.37\\
    \small{LCC-D\tnote{*}}&  0.79 $\pm$ N/A &  9.21 &N/A&&-&-&-\\
    \small{VM\tnote{*}}  &  0.79 $\pm$ N/A & 8.46 & N/A& &-&-&-\\
    \small{LVM (S1)\tnote{*}} &  0.79 $\pm$ N/A &  7.58 &N/A&&-&-&-\\
    \small{SyN\tnote{*}}&  0.80 $\pm$ N/A &  8.24 & N/A&& 0.81 $\pm$ 0.16 &  8.9 & 0.02\\
    \small{MM}& 0.81 $\pm$ 0.18 &  8.59 & 0 && 0.72 $\pm$ 0.12 &  3.48 & 0 \\
    \small{LVM} (S3)\tnote{*} &  0.81 $\pm$ N/A &  6.88 & N/A &&-&-&-\\
    Demons &-&-&-&& 0.65 $\pm$ 0.18 & 15.46 & 0.42 \\
    DIRNet\tnote{**} &-&-&-& & 0.80 $\pm$ 0.08 &  N/A & N/A\\
    \setrow{\bfseries}Ours &\setrow{\bfseries}  0.92 $\pm$ 0.10 &\setrow{\bfseries}  4.76 &\setrow{\bfseries} 0 & &\setrow{\bfseries} 0.85 $\pm$ 0.15 &\setrow{\bfseries}  5.06 &\setrow{\bfseries} 0\\
    \bottomrule
  \end{tabularx}
  \end{threeparttable}
  \end{minipage}%
  
  \begin{minipage}[b]{1\linewidth}
\captionsetup{justification=centering}
  \centering
\begin{threeparttable}
 \begin{tabularx}{1\linewidth}{ lcccccccccc }
    
    Methods &\multicolumn{2}{X}{}&LA&\multicolumn{6}{X}{}\\
    \cmidrule(lr){2-11}
         & \multicolumn{3}{c}{Dice} & \multicolumn{3}{c}{HD}  & \multicolumn{3}{r}{$\det(J_{\phi})<0$}&\\
 
    \cmidrule(lr){2-4} \cmidrule(lr){5-7} \cmidrule(lr){8-11}
     & 2ch & 3ch & 4ch & 2ch & 3ch & 4ch &&2ch & 3ch & 4ch  \\
    \midrule
    Unregistered  &  0.79 $\pm$ 0.07 &  0.78 $\pm$ 0.08 &  0.78 $\pm$ 0.09 & 7.37 & 7.70 & 8.66 && N/A & N/A & N/A \\
    Elastix &  0.81 $\pm$ 0.13 &  0.86 $\pm$ 0.08 &  0.80 $\pm$ 0.11 & 7.9 & 6.99 & 8.01 && 0.68 & 0.5 & 0.75\\
    RTT &  0.82 $\pm$ 0.11&  0.85 $\pm$ 0.10&  0.80 $\pm$ 0.10 & 7.28 & 6.82 & 7.56 && 0.53 & 0.45 & 0.7\\
    MM &  0.84 $\pm$ 0.06 &  0.83 $\pm$ 0.06 &  0.83 $\pm$ 0.08 & 6.58 & 6.48 & 6.77  && 0.01 & 0 & 0 \\
    Demons &  0.84 $\pm$ 0.08 &  0.85 $\pm$ 0.06 &  0.82 $\pm$ 0.10 & 7.41 & 7.33 & 7.84  && 0.025 & 0.042 & 0\\
    SyN &  0.87 $\pm$ 0.06&  0.86 $\pm$ 0.13&  0.84 $\pm$ 0.11 & 6.92 & 7.52 & 7.51 & &0.033 & 0 & 0.01 \\
    \setrow{\bfseries}Ours &\setrow{\bfseries}  0.89 $\pm$ 0.11 & \setrow{\bfseries}  0.92 $\pm$ 0.06 &\setrow{\bfseries}  0.86 $\pm$ 0.11 & \setrow{\bfseries}  6.26 &\setrow{\bfseries}  5.7 &\setrow{\bfseries} 7.44 & &0 & 0 & 0\\

    \bottomrule
 \end{tabularx}
 \begin{tablenotes}
  \item[]$^{*}$Results are are reported in \cite{krebs2019learning}. $^{**}$Results are reported in \cite{de2017end}. N/A refers to the value that are not reported in the original paper.
  \end{tablenotes}
  \end{threeparttable}
  \end{minipage}%
  \label{table:leftacdcscd}
  \vspace{-6mm}
\end{table}

\section{Experiments}
\subsubsection{Data:} The proposed framework is evaluated on three clinical cardiac MRI datasets: 1) Automated Cardiac Diagnosis Challenge (ACDC) dataset~\cite{bernard2018deep} which contains MRI scans including 100 stack of short-axis cardiac cine MRI sequences, each consisting of 12 to 35 temporal frames. 2) SunnyBrook cardiac dataset (SCD) \cite{radau2009evaluation} which contains 45 cardiac cine MRI scans each with 20 temporal frames. 3) Left atrium (LA) from the 2-chamber (2ch), 3-chamber (3ch) and 4-chamber (4ch) cine long-axis MRI sequences acquired retrospectively at our center that consists of 120 pairs of images in each chamber group. In all three datasets, we applied the proposed framework on end-diastole (ED) and end-systole (ES) cardiac frames from the cine sequences.

\subsubsection{Implementation:} The proposed method is implemented in Python programming language using Pytorch module. In neural network design, the Adam optimization with $800$ update iterations and a learning rate of $5\times10^{-4}$  are used for all the three datasets. The level of pyramid $K$ is considered $2$ with $\alpha = 1/(K)$, $\gamma =\frac{\lambda\times1}{K}$, $\lambda = 5$. The neural net framework is evaluated on an NVIDIA GeForce GTX 1080 Ti GPU.
\subsubsection{Evaluation Metrics:}
The Dice metric (DM), reliability, Hausdorff distance (HD) and the determinant of Jacobian are used as the evaluation metrics to compare the performance of the algorithms.

\paragraph{Dice Metric}
The DM \cite{dice1945measures} is a well-known segmentation based metric to measure the similarity (overlap) between two regions, warped moving and fixed image. The DM of two regions A and B is formulated as:
\begin{equation}
    \label{eq:Dice}
        DM(A,B) = \frac{2 |A \cap B|}{A+B}
\end{equation}%
\paragraph{Reliability}
The algorithms are examined by evaluating the reliability function \cite{ayed2009left} of the obtained DMs using \eqref{eq:relability}. The complementary cumulative distribution function is defined for each $d \in [0, 1]$ as the probability of obtaining $DM$ higher than $d$ over the entire set. The $R(d)$ measures how reliable is the algorithm in yielding accuracy $d$.
\begin{equation}
    \label{eq:relability}
    \begin{split}
        R(d) &= P_r(Dice > d)
        = \frac{\textnormal{\# Images segmented with DM higher than d}}{\textnormal{total number of images}}
    \end{split}
\end{equation}%
\paragraph{Hausdorff Distance}
The HD \cite{huttenlocher1993comparing} is another well-known metric which measures the maximum deviation between two regions' contours. The HD between two contours $(C_A)$ and $C_B$ is formulated as:
\begin{equation}
    \label{eq:HD}
        \textnormal{HD}(C_A,C_B) = \max(\max_{i}(\min_{j}(d(p^{i}_{A},p^{j}_{B}))),\max_{j}(\min_{i}(d(p^{i}_{A},p^{j}_{B}))))
\end{equation}%
where $p^{i}_{A}, p^{j}_{B}$ denote the set of all the points in $C_A$ and $C_B$ respectively. The term $d(\cdot)$ denotes the Euclidean distance.
\paragraph{Determinant of Jacobian}
We quantify and analyze deformation regularity using the determinant of the Jacobian, $\det(J_{\phi})$. Jacobian matrix captures the local behaviours of the deformation field. If the value of $\det(J)$ equals to $1$, the volume remains constant after the transformation, the value smaller than $1$ shows the local volume shrinkage and greater than $1$ shows the local volume expansion. The negative values imply that local folding and twists have occurred, which is physically not realizable and mathematically not inevitable \cite{dalca2018unsupervised}.

\begin{table}[!t]
\caption{Quantitative cardiac MRI registration results on $50\%$ patients of ACDC, SCD and LA datasets based on different loss functions. The evaluation was performed in terms of average Dice, HD and the forward and backward registration time (per second).}
    \begin{minipage}{.5\linewidth}
    \centering
        \begin{tabular}{cccc}
         &SCD&&\\
        \toprule
        \small{Methods} &\small{ Dice} & \small{HD }& \small{Time(s)}\\
        \cmidrule(lr){2-2}
        \cmidrule(lr){3-3}
        \cmidrule(lr){4-4}
        \small{MSE} & \small{0.83 } & \small{ 9.1} &\small{15}\\
        \small{SSIM} &\small{0.86 } &\small{ 7.9 } &\small{20 }\\
        \small{SSIM + MI} &\small{0.89 } &\small{6.01} &\small{26} \\
        \bottomrule
        \end{tabular}
    \end{minipage}%
    \begin{minipage}{0.5\linewidth}
    \centering
        \begin{tabular}{cccc}
         &ACDC&&\\
        \toprule
        \small{Methods} &\small{ Dice} & \small{HD }& \small{Time(s)}\\
        \cmidrule(lr){2-2}
        \cmidrule(lr){3-3}
        \cmidrule(lr){4-4}
        \small{MSE} & \small{0.87} & \small{5.73 } &\small{15}\\
        \small{SSIM} & \small{0.92 }& \small{4.31}&\small{21} \\
        \small{SSIM + MI} &\small{0.93} &\small{4.31} &\small{35} \\
        \bottomrule
        \end{tabular}%
    \end{minipage}
    
    \begin{minipage}{1\linewidth}
    \centering
        \begin{tabularx}{1\linewidth}{ cXXXXXXc }
        \centering
        &&&&&&&\\
        &&&&LA&&&\\
        \toprule
            Methods & \multicolumn{3}{c}{Dice} & \multicolumn{3}{c}{HD} & Time (s)\\
        \cmidrule(lr){2-4}
        \cmidrule(lr){5-7}
        \cmidrule(lr){8-8}
         & 2ch & 3ch & 4ch & 2ch & 3ch & 4ch\\
        \cmidrule(lr){2-4}
        \cmidrule(lr){5-7}
        \cmidrule(lr){8-8}
        \small{MSE} &  \small{0.84 } &  \small{0.86} &  \small{0.84} & \small{6.83} & \small{6.88} & \small{7.84 } &\small{15}\\
        \small{SSIM} &  \small{0.91} &  \small{0.93 }& \small{0.89} &\small{ 6.26} & \small{6.11} & \small{6.42 }& \small{22}\\
        \small{SSIM + MI} &  \small{0.90} &   \small{0.93}  &\small{0.87} & \small{6.89} & \small{5.22}& \small{6.40} & \small{35}\\
        \bottomrule
        \end{tabularx}
    \end{minipage}%
\label{tale:ACDCscdleftlosses}
\end{table}%
\begin{figure*}[!t]
	  \setlength{\belowcaptionskip}{-8pt}
		\begin{minipage}[b]{0.325\linewidth}
		\centering

		\centerline{\includegraphics[width=\textwidth]{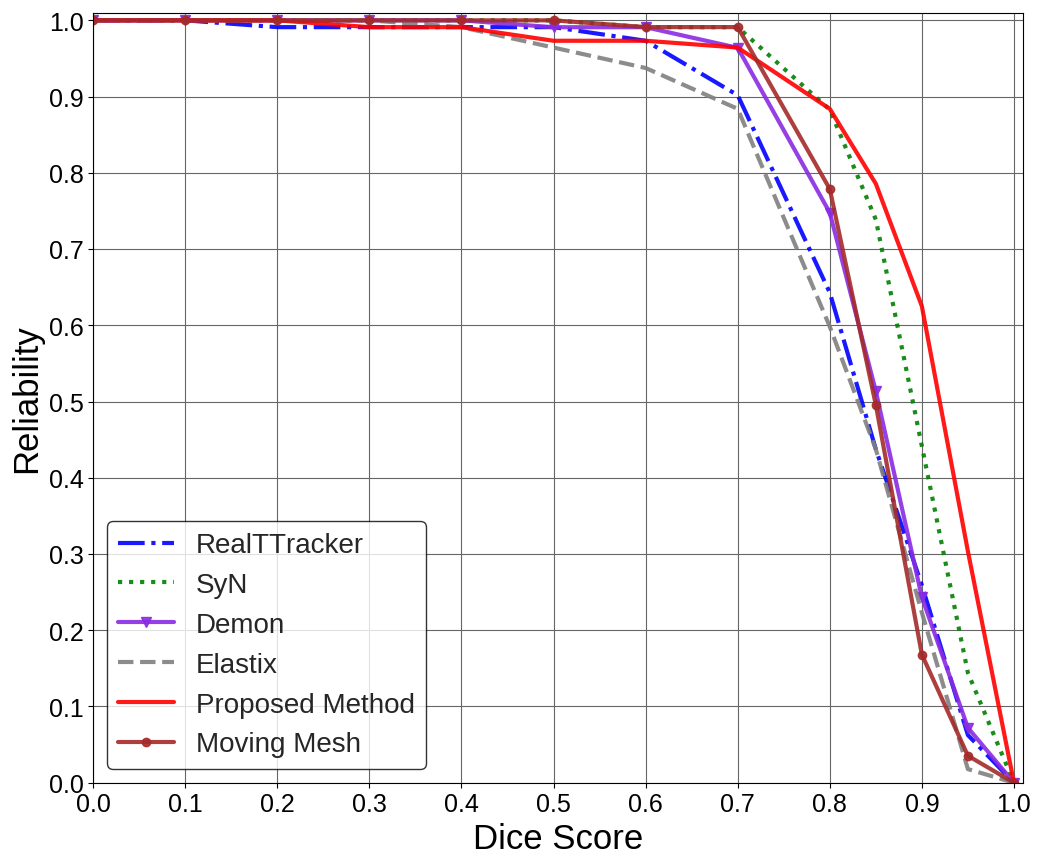}}
		\centerline{2ch}\medskip
	\end{minipage}
	\hfill
	\begin{minipage}[b]{0.325\linewidth}
		\centering

		\centerline{\includegraphics[width=\textwidth]{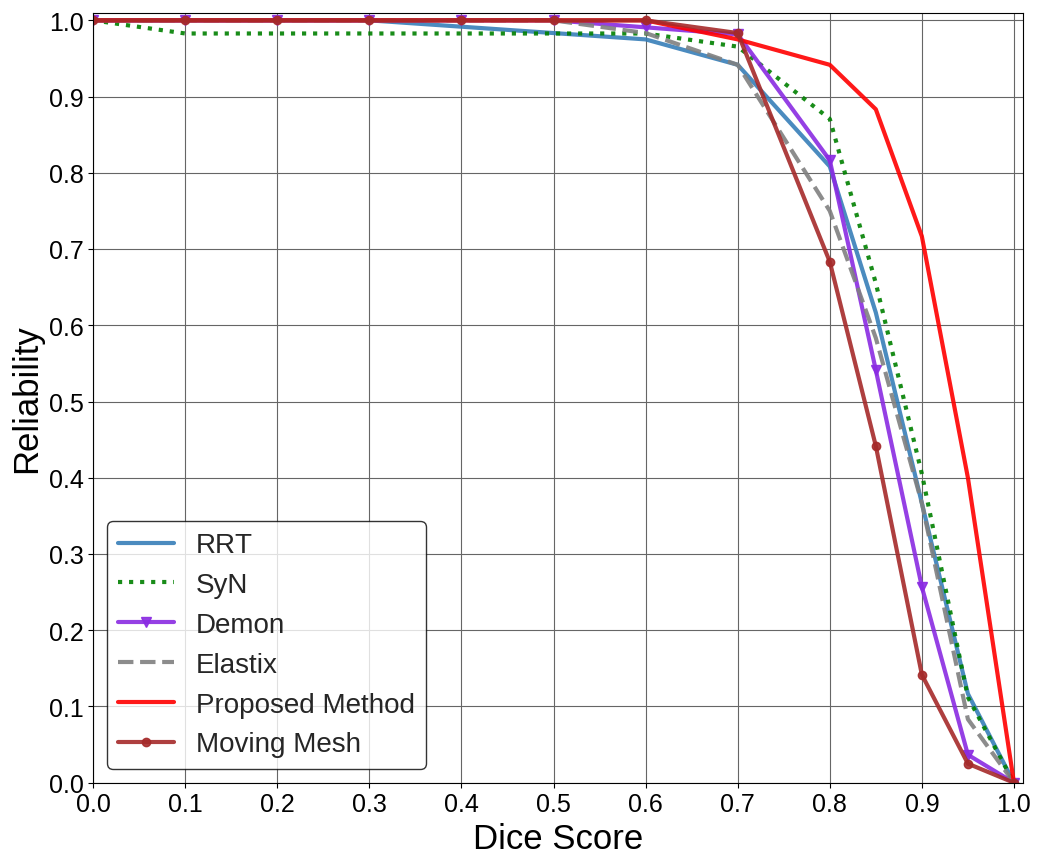}}
		\centerline{3ch}\medskip 
	\end{minipage}
	\hfill
	\begin{minipage}[b]{0.325\linewidth}
		\centering

		\centerline{\includegraphics[width=\textwidth]{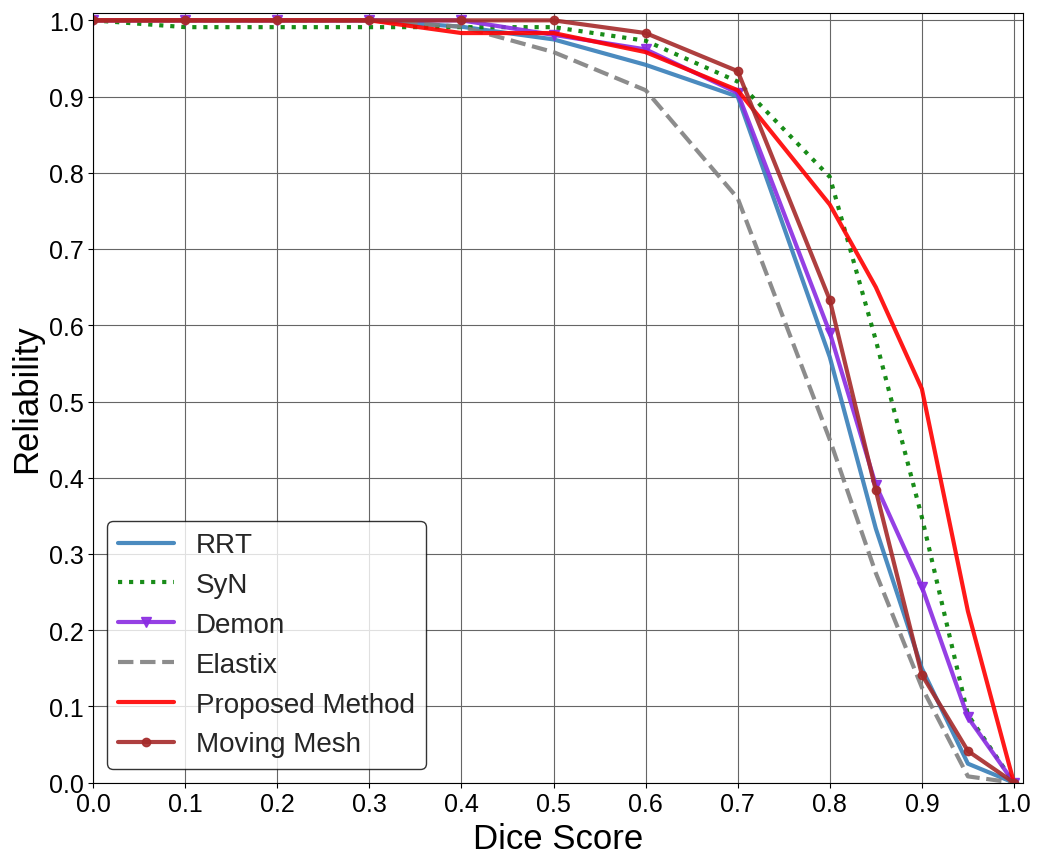}}
		\centerline{4ch}\medskip
	\end{minipage}
	\begin{minipage}[b]{1\linewidth}
		\centering
		\centerline{Left Atrium}\medskip 
	\end{minipage}
	
	\begin{minipage}[b]{0.45\linewidth}
		\centering

		\centerline{\includegraphics[width=0.72\textwidth]{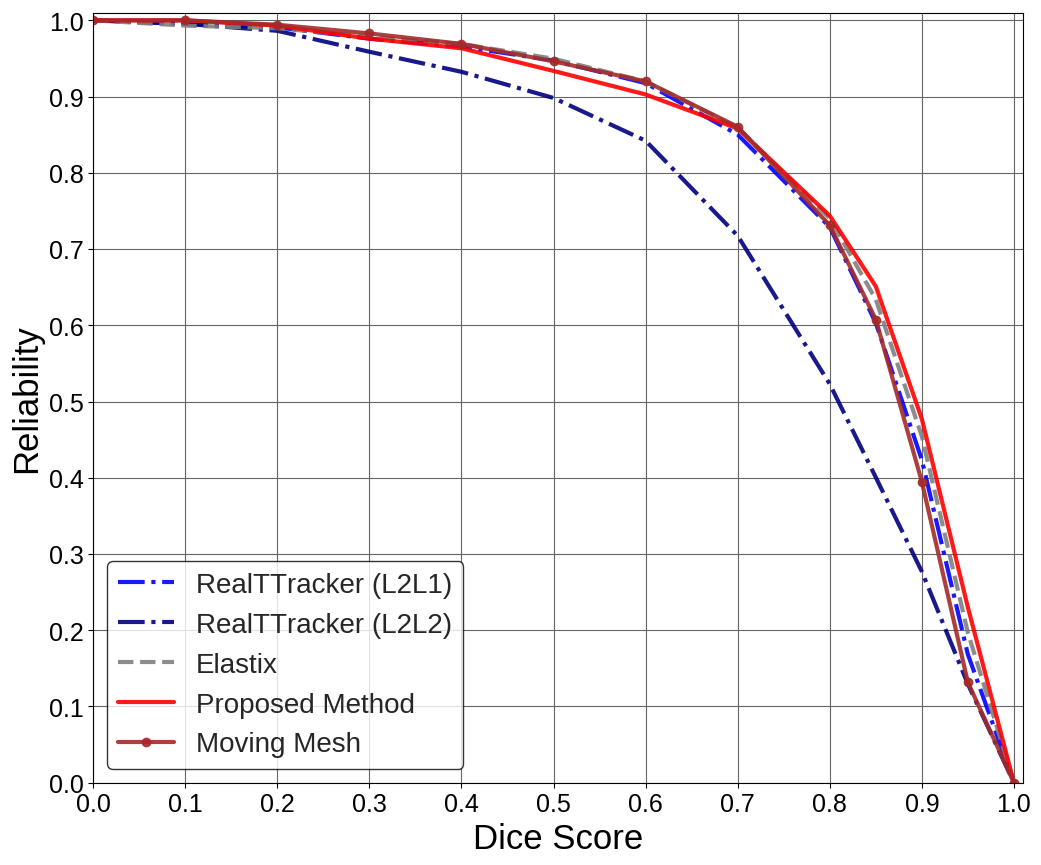}}
		\centerline{ACDC}\medskip 
	\end{minipage}
	\hfill
	\begin{minipage}[b]{0.45\linewidth}
		\centering

		\centerline{\includegraphics[width=0.72\textwidth]{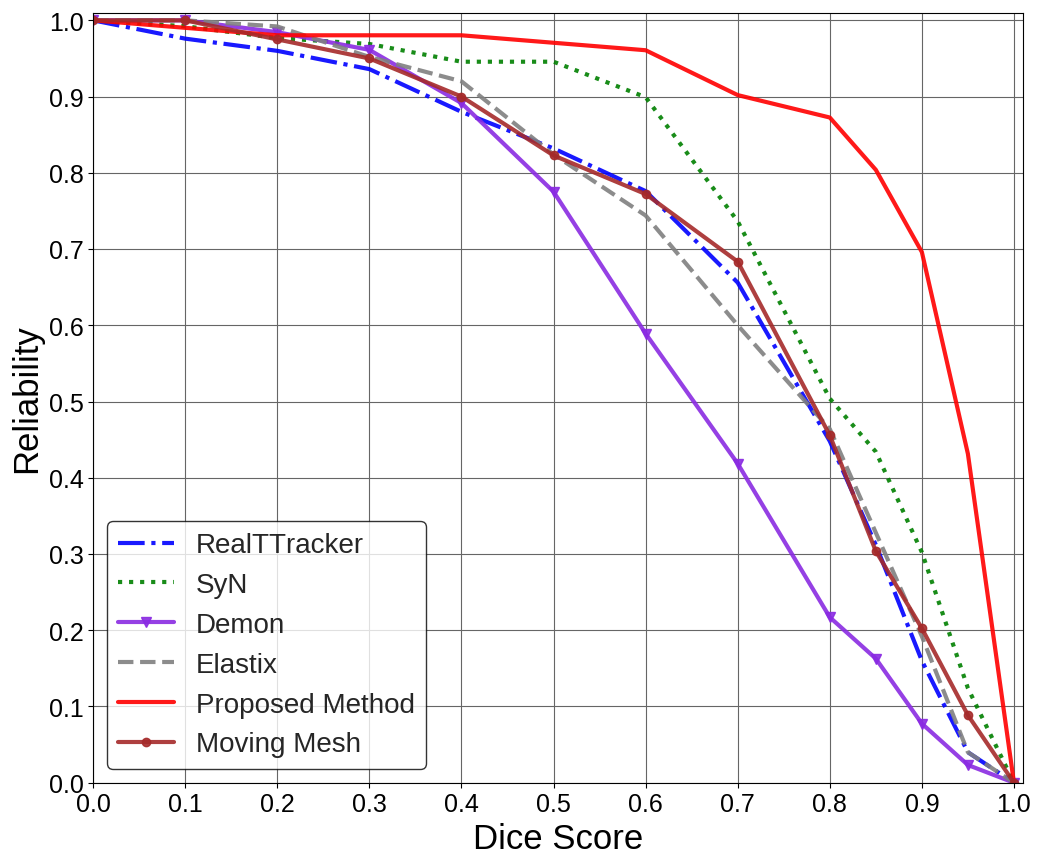}}
		\centerline{SCD}\medskip
	\end{minipage}
	\caption{Reliability versus Dice metric for different algorithms. The figures omit results for methods with no public implementation available.} 
	\label{fig:reliabilirty}
\end{figure*}%

\begin{table}[!th]
 \centering
\begin{minipage}[b]{0.8\linewidth}
\captionsetup{justification=centering}
  \caption{Quantitative cardiac MRI registration results on the ACDC, SCD and LA based on different number of resolution. The evaluation is performed in terms of Dice (mean$\pm$ standard deviation) and HD (mean). The 2ch, 3ch and 4ch stand for the 2, 3 and 4-chamber. Values in bold indicate the best performance.}
  \centering
 \begin{tabularx}{\linewidth}{llccc}
    \toprule
    \centering
       & Dataset & & Dice & HD\\
    \cmidrule(lr){2-3}
    \cmidrule(lr){4-5}
    \multirow{4}{6em}{One Resolution} & ACDC & & 0.85 $\pm$ 0.20  &  6.03 \\
    & SCD& & 0.85 $\pm$ 0.17  & 7.70\\
    & LA-2ch & & 0.87 $\pm$ 0.08 & 7.6\\
    & LA-3ch & & 0.91 $\pm$ 0.05 & 6.59\\
    & LA- & & 0.88 $\pm$ 0.10 & 6.63\\
    \midrule
    \multirow{4}{6em}{\setrow{\bfseries}Two Resolutions} & ACDC & & \setrow{\bfseries}0.87 $\pm$ 0.17  &\setrow{\bfseries} 6.27 \\
    & SCD & & \setrow{\bfseries}0.86 $\pm$ 0.16  & \setrow{\bfseries}7.06\\
    & LA-2ch & & \setrow{\bfseries}0.89 $\pm$ 0.09 & \setrow{\bfseries}7.59\\
    & LA-3ch & &\setrow{\bfseries} 0.93 $\pm$ 0.03 & \setrow{\bfseries}5.47\\
    & LA-4ch & & \setrow{\bfseries}0.87 $\pm$ 0.11 & \setrow{\bfseries}6.67\\
    \midrule
    \multirow{4}{6em}{Three Resolutions} & ACDC & &0.83 $\pm$ 0.22 & 6.70 \\
    & SCD & & 0.85 $\pm$ 0.18 & 7.22 \\
    & LA-2ch & & 0.86 $\pm$ 0.09 &  7.03\\
    & LA-3ch & & 0.90 $\pm$ 0.06 & 6.22\\
    & LA-4ch & & 0.86 $\pm$ 0.11 & 6.42\\
    \midrule
    \multirow{4}{6em}{Four Resolutions} & ACDC & & 0.74 $\pm$ 0.24  & 8.96 \\
    & SCD & & 0.85 $\pm$ 0.19  & 7.50\\
    & LA-2ch & & 0.82 $\pm$ 0.10 & 7.06\\
    & LA-3ch & & 0.86 $\pm$ 0.07 & 6.93\\
    & LA-4ch & & 0.81 $\pm$ 0.10 & 7.01\\
    \bottomrule
\end{tabularx}
  \label{table:resoulotions}
  \end{minipage}%
\end{table}
\subsection{Evaluation Results and Discussions}
The performance of the proposed framework is compared with nine state-of-the-art learning-based and classic deformable registration algorithms including SimpleElastix (Elastix) \cite{marstal2016simpleelastix}, Moving Mesh (MM) \cite{punithakumar2017gpu}, Real-Time Image-based Tracker (RTT) \cite{zachiu2015improved}, Demons \cite{mccormick2014itk}, LCC-Demons (LCC-D) \cite{lorenzi2013lcc}, Symmetric Normalization (SYN) \cite{avants2008symmetric}, \cite{dalca2018unsupervised}, \cite{krebs2019learning} and DIRNet \cite{de2017end}. 

\begin{figure*}
\begin{minipage}[b][1cm]{0.32\linewidth}
		\centering
		\centerline{\includegraphics[ width=\textwidth]{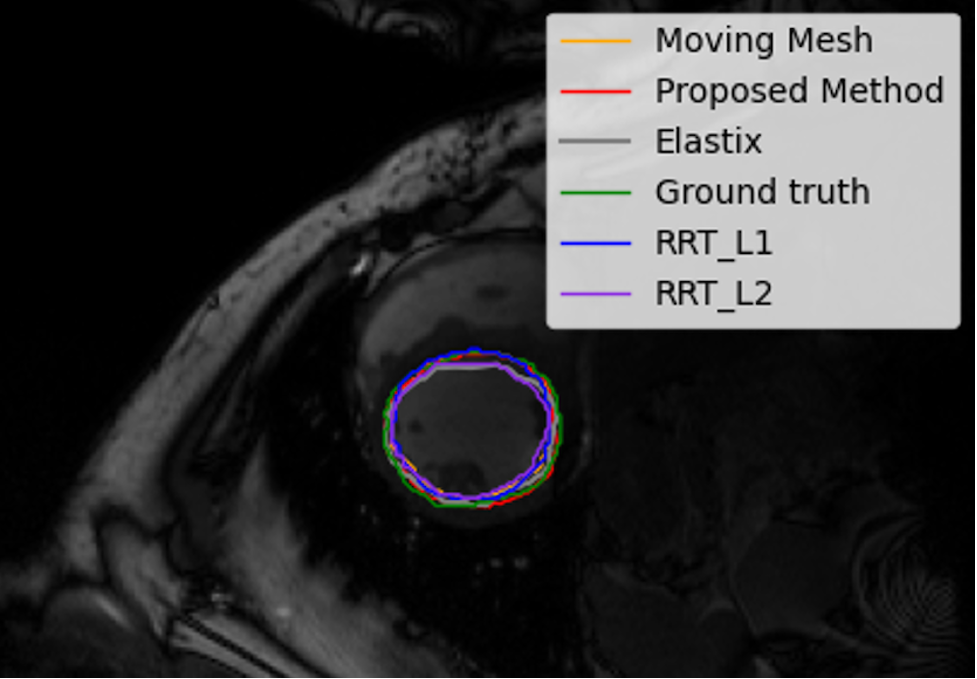}}
	\end{minipage}
	\hfill
	\begin{minipage}[b]{0.32\linewidth}
		\centering
		\centerline{\includegraphics[ width=\textwidth]{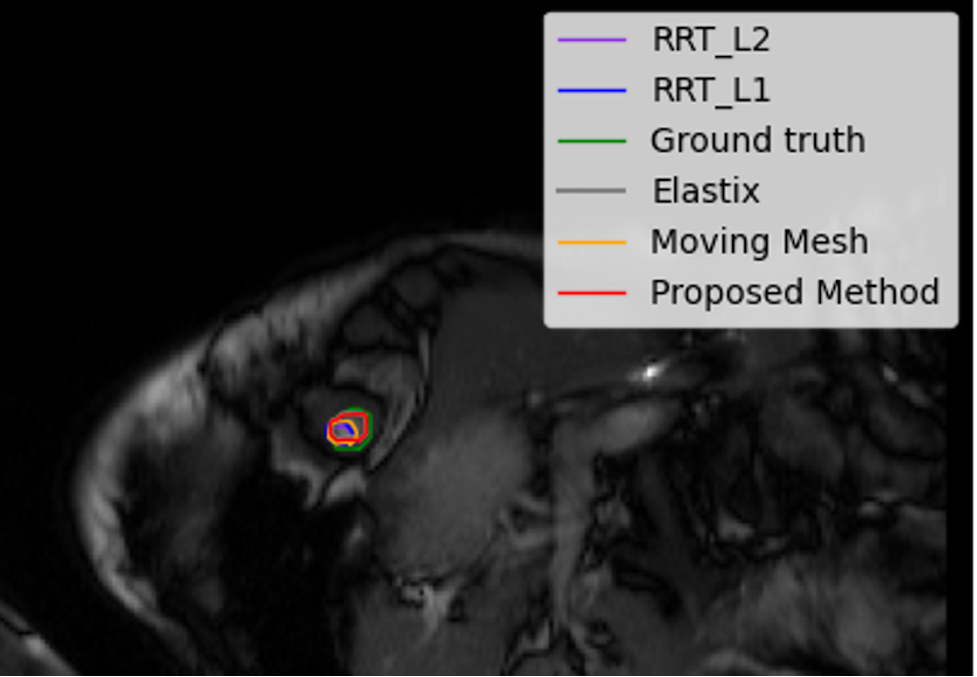}}
	\end{minipage}
	\hfill
	\begin{minipage}[b]{0.32\linewidth}
		\centering
		
		\centerline{\includegraphics[ width=\textwidth]{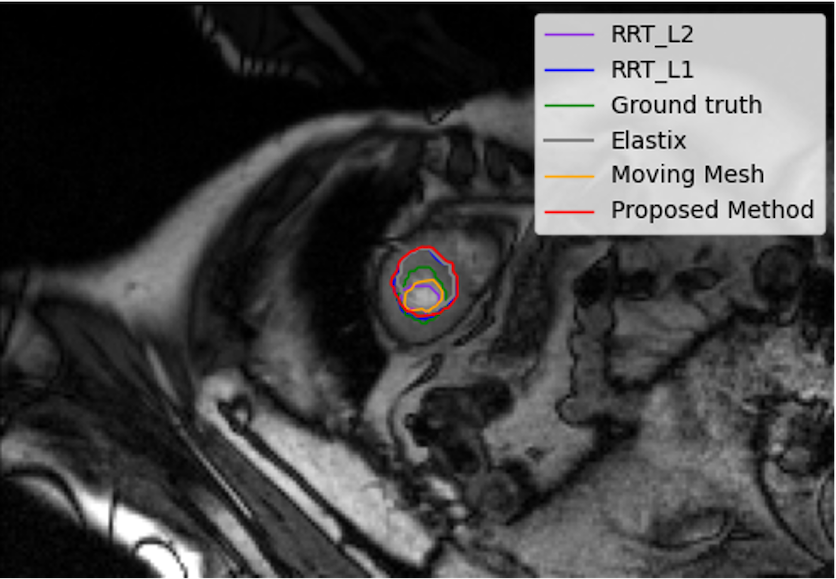}}
	\end{minipage}
	\begin{minipage}[b]{\linewidth}
		\centering
	
		\centerline{ACDC}\medskip 
	\end{minipage}
	\begin{minipage}[b][1cm]{0.32\linewidth}
		\centering
		\centerline{\includegraphics[ trim={0cm 2cm 0cm 0cm},clip,width=\textwidth]{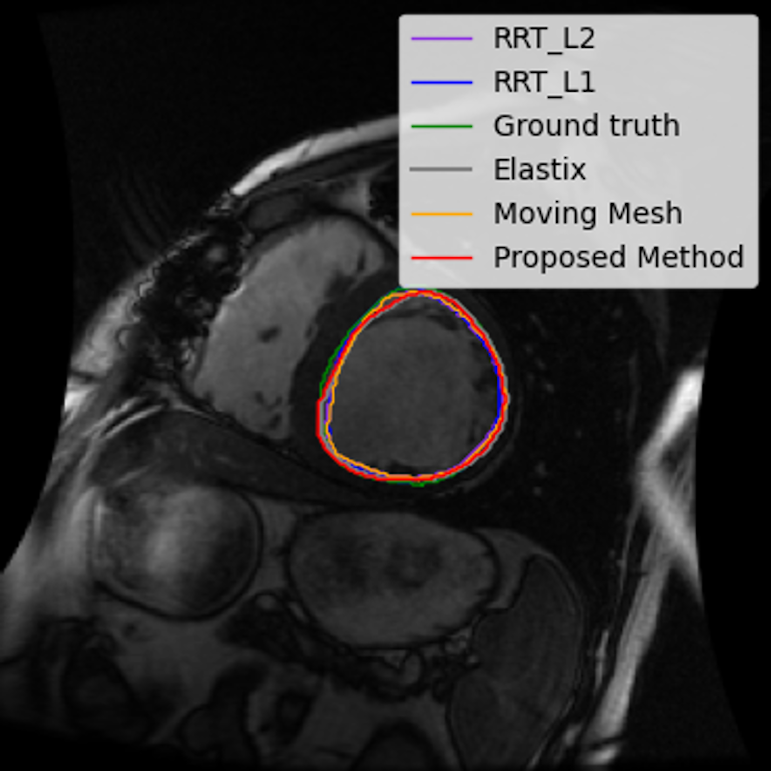}}
	\end{minipage}
	\hfill
	\begin{minipage}[b]{0.32\linewidth}
		\centering
		\centerline{\includegraphics[ trim={0cm 2cm 0cm 0cm},clip,width=\textwidth]{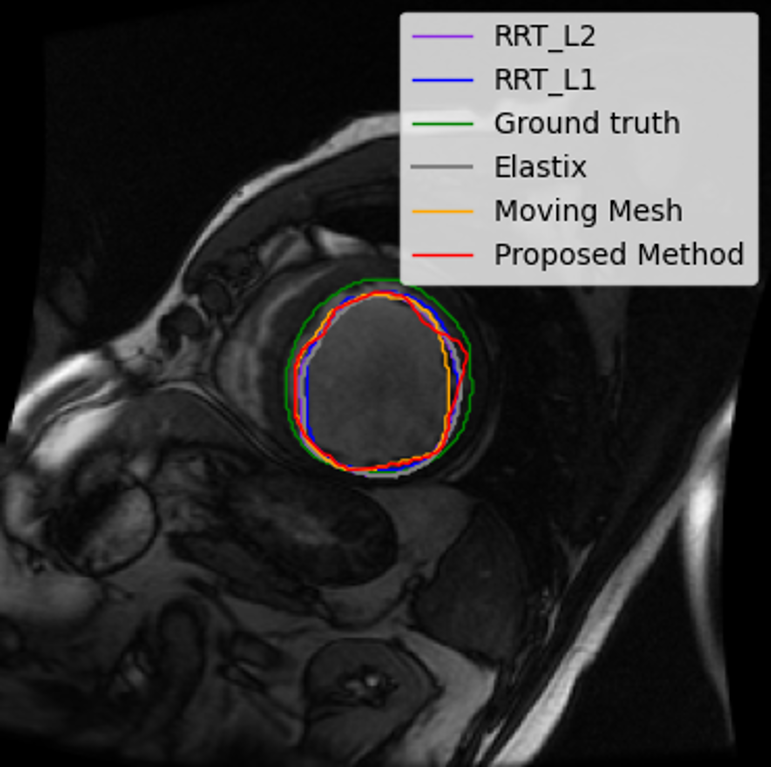}}
	\end{minipage}
	\hfill
	\begin{minipage}[b]{0.32\linewidth}
		\centering
		
		\centerline{\includegraphics[trim={0cm 2cm 0cm 0cm},clip, width=\textwidth]{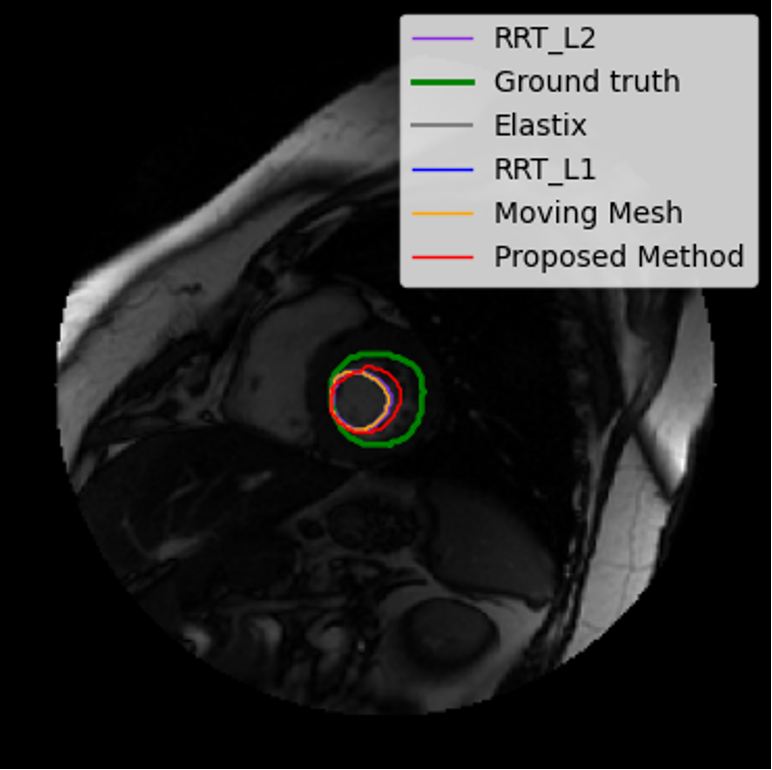}}
	\end{minipage}
	\begin{minipage}[b]{\linewidth}
		\centering
	
		\centerline{SGD}\medskip 
	\end{minipage}
	\begin{minipage}[b][1cm]{0.32\linewidth}
		\centering
		\centerline{\includegraphics[ width=\textwidth]{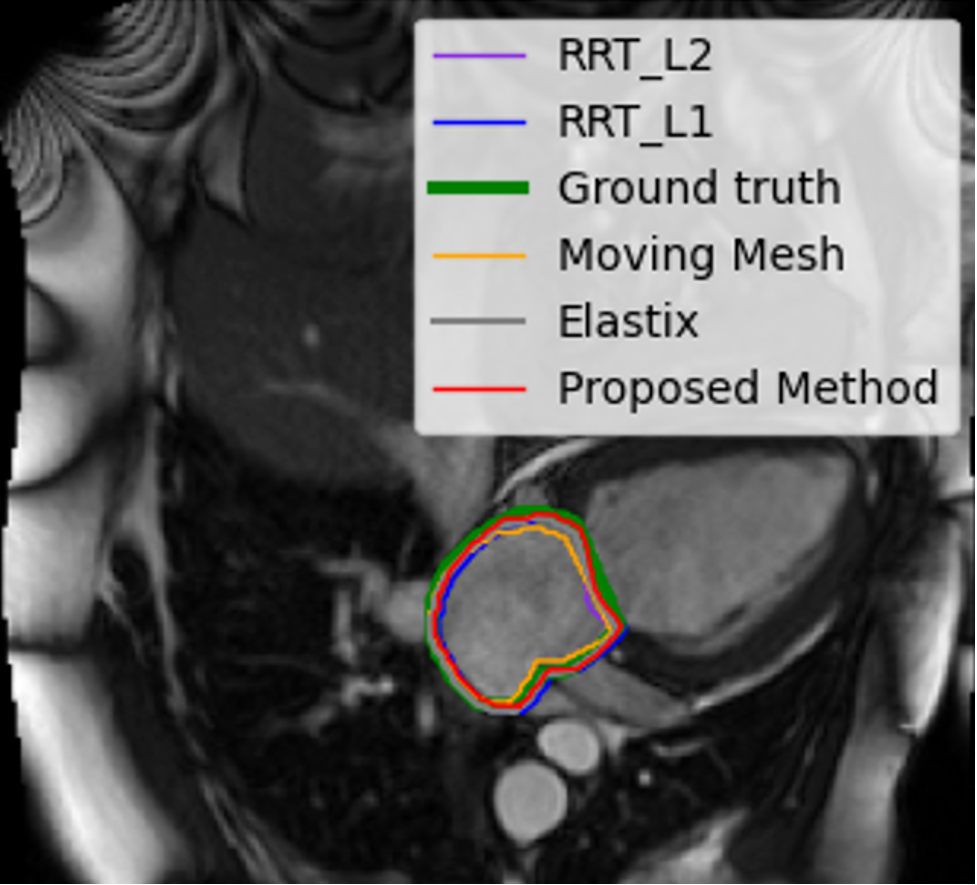}}
	\end{minipage}
	\hfill
	\begin{minipage}[b]{0.32\linewidth}
		\centering
		\centerline{\includegraphics[ width=\textwidth]{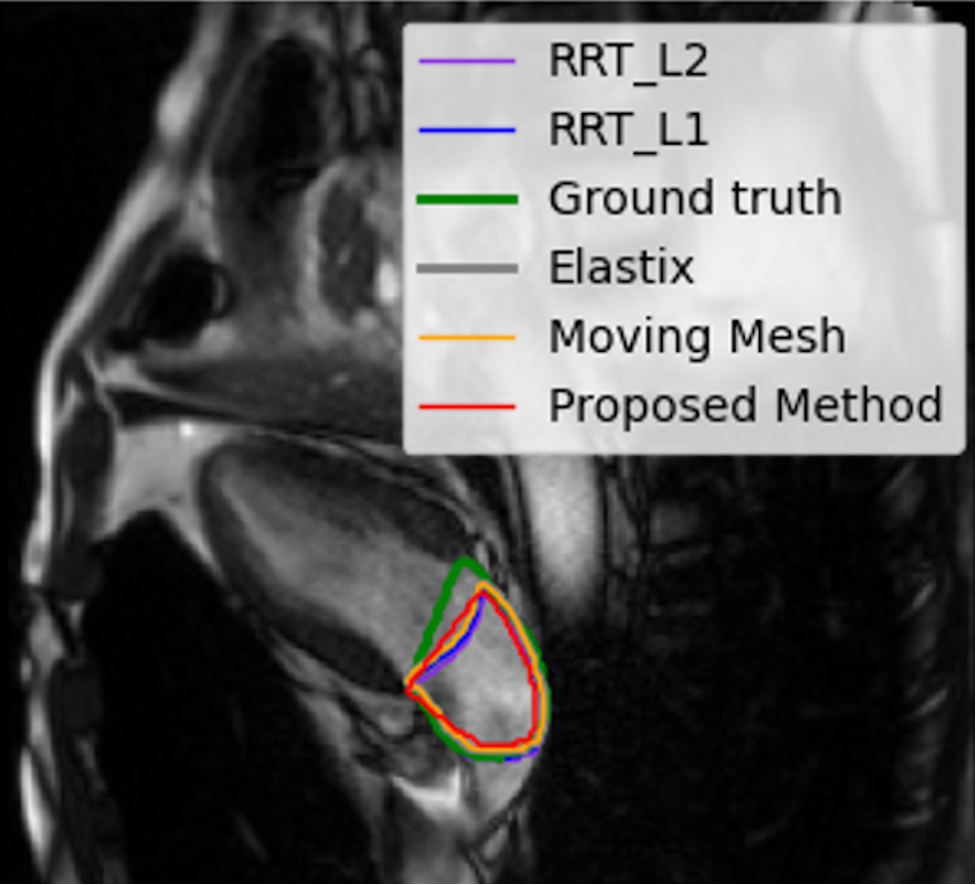}}
	\end{minipage}
	\hfill
	\begin{minipage}[b]{0.32\linewidth}
		\centering
		
		\centerline{\includegraphics[trim={0cm 0.3cm 0cm 0cm},clip, width=\textwidth]{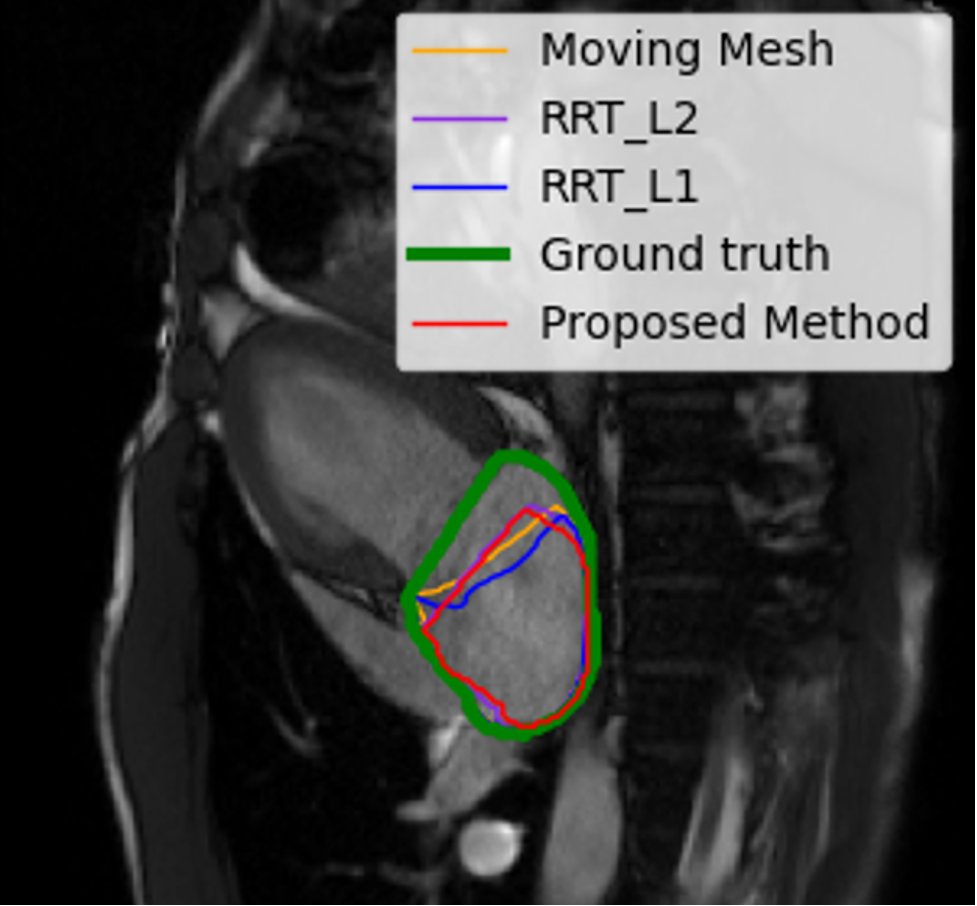}}
	\end{minipage}
	\begin{minipage}[b]{\linewidth}
		\centering
	
		\centerline{Left-2ch}\medskip 
	\end{minipage}
		\begin{minipage}[b][1cm]{0.32\linewidth}
		\centering
		\centerline{\includegraphics[trim={0cm 1cm 0cm 0cm},clip, width=\textwidth]{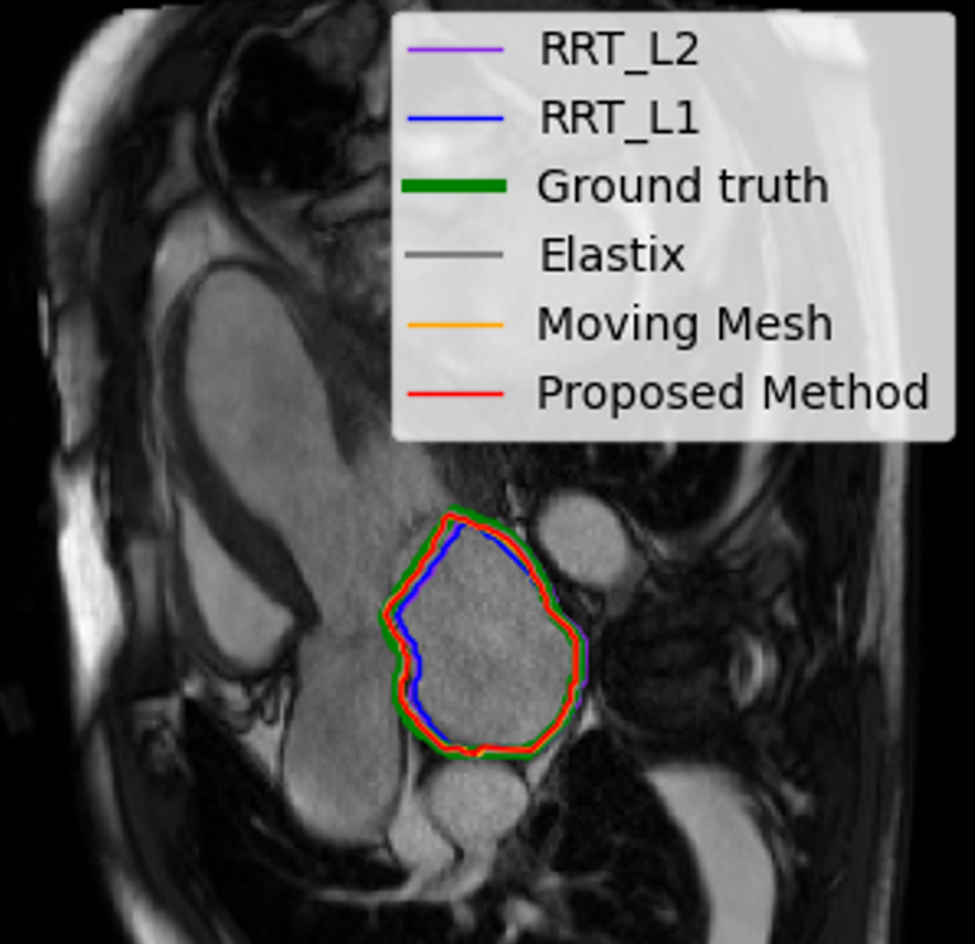}}
	\end{minipage}
	\hfill
	\begin{minipage}[b]{0.32\linewidth}
		\centering
		\centerline{\includegraphics[ width=\textwidth]{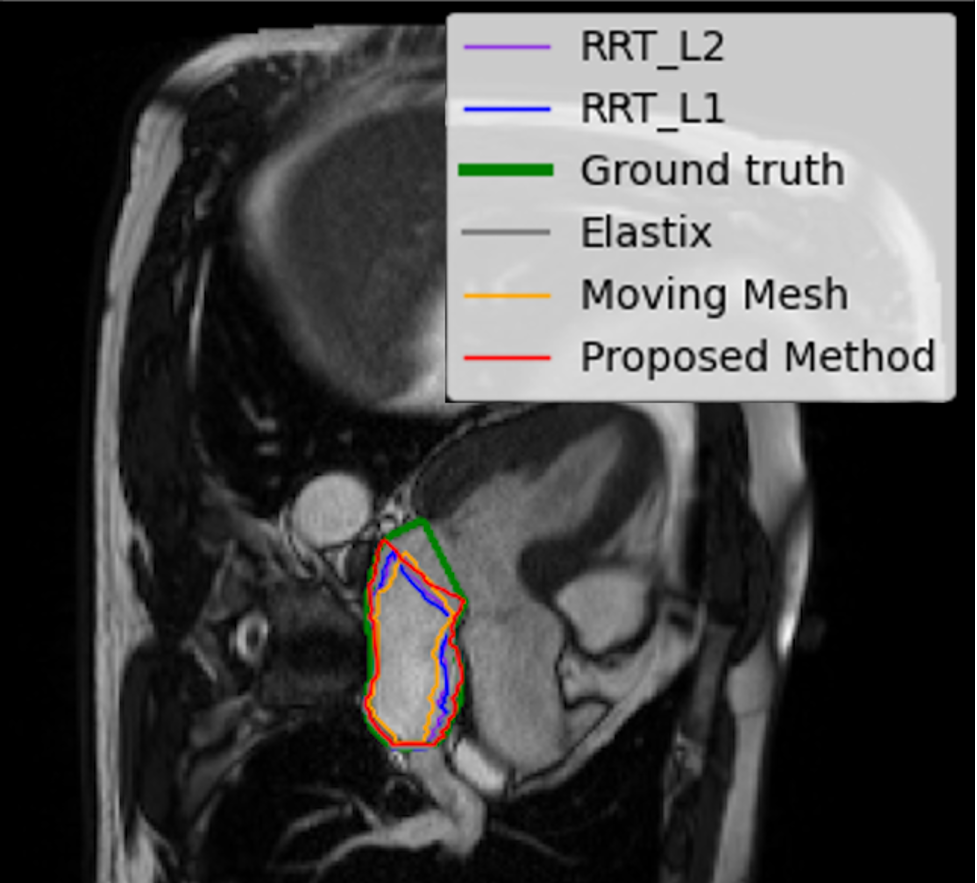}}
	\end{minipage}
	\hfill
	\begin{minipage}[b]{0.32\linewidth}
		\centering
		
		\centerline{\includegraphics[trim={0cm 0.8cm 0cm 0cm},clip, width=\textwidth]{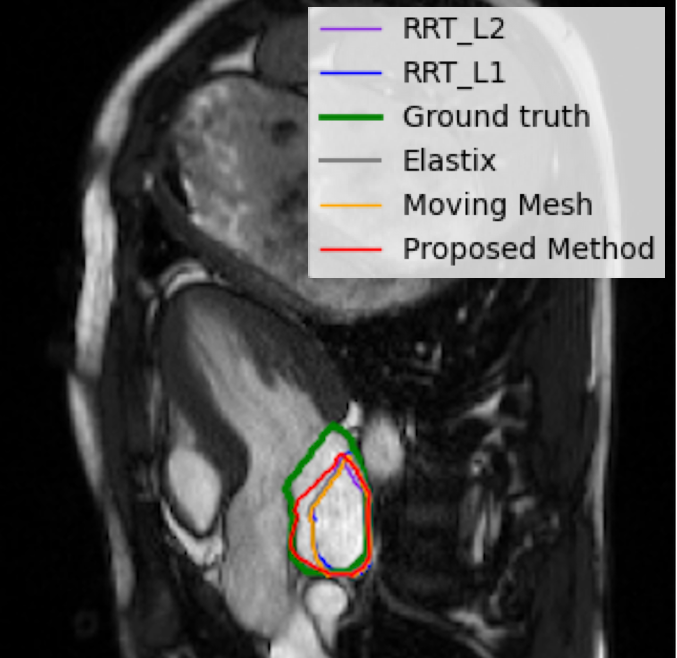}}
	\end{minipage}
	\begin{minipage}[b]{\linewidth}
		\centering
	
		\centerline{Left-3ch}\medskip 
	\end{minipage}
		\begin{minipage}[b][1cm]{0.32\linewidth}
		\centering
		\centerline{\includegraphics[ width=\textwidth]{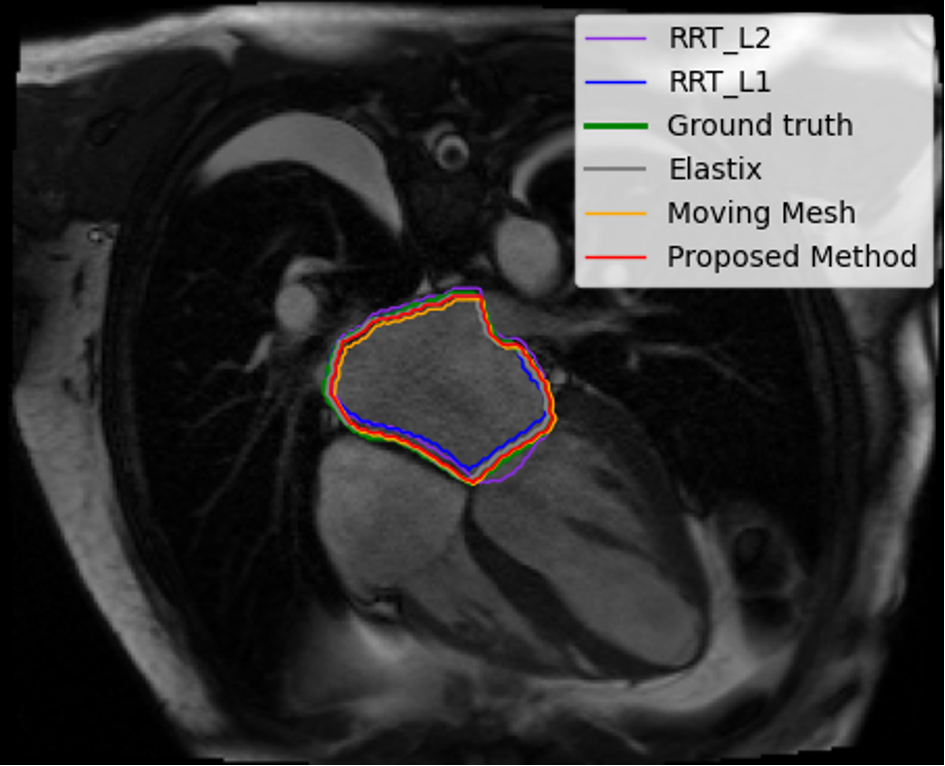}}
	\end{minipage}
	\hfill
	\begin{minipage}[b]{0.32\linewidth}
		\centering
		\centerline{\includegraphics[trim={0cm 0.9cm 0cm 0cm},clip, width=\textwidth]{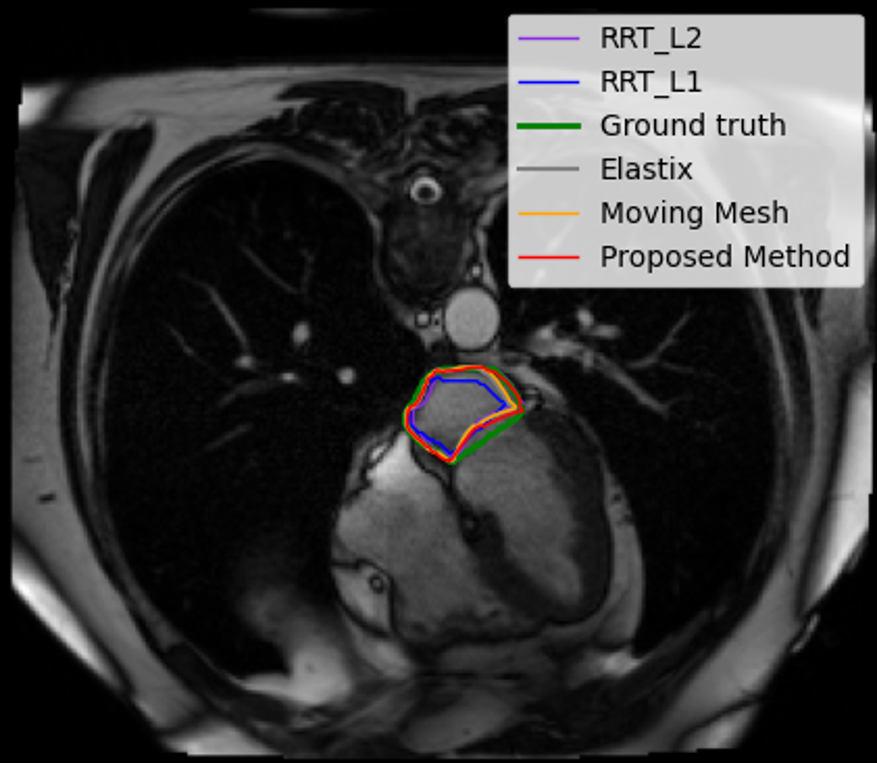}}
	\end{minipage}
	\hfill
	\begin{minipage}[b]{0.32\linewidth}
		\centering
		
		\centerline{\includegraphics[ width=\textwidth]{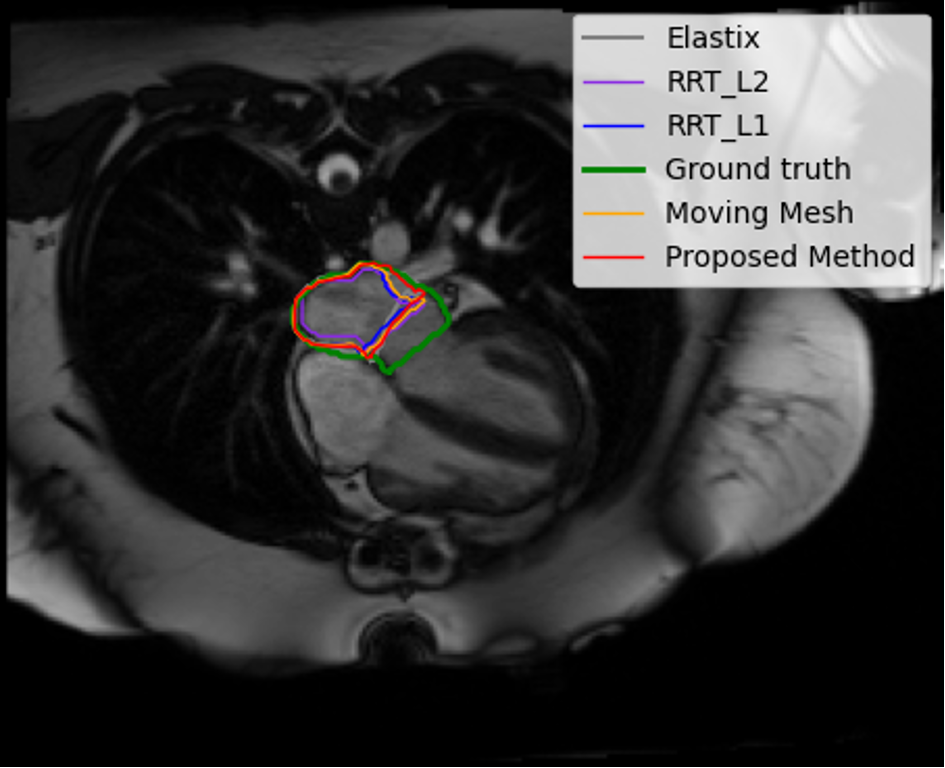}}
	\end{minipage}
	\begin{minipage}[b]{\linewidth}
		\centering
	
		\centerline{Left-4ch}\medskip 
	\end{minipage}

	\caption{The segmentation results with the highest (first column), 50th percentile (2nd column), lowest (3rd column) Dice score values for (a) ACDC (1st row), (b) SGD (2nd row), and left atrium (c) 2-chamber (3rd row), (d) 3-chamber (4th row), 4-chamber (5th row) views. The green represent the ground truth, also the red, blue, violet, gray and orange contours represent the boundary corresponds to the registration by proposed method, RRT1, RRT2, Elastix and MM, respectively}
	\label{fig:precentile}
	\vspace{-0.2cm}
\end{figure*}

Table~\ref{table:leftacdcscd}, shows the mean and standard deviations of Dice scores, HD and the average number of pixels with non-positive $\det(J_{\phi})$ over all subjects ACDC, SCD, and left atrium datasets. Compared to the conventional and learning-based methods, our framework has the best performances in terms of Dice scores. In Fig. \ref{fig:precentile}, the highest accuracy, $50th$ percentile, and the lowest accuracy in terms of Dice scores achieved by the proposed method are shown visually, where the red, blue, violet, gray and orange contours depict the correspondence segmentation results by the proposed method, RRT1, RRT2, Elastix and MM, respectively and the green contours depict the ground truth.

To quantify and analyse deformation regularity, we computed the determinant of the Jacobian $\det(J_{\phi})$ using SimpleITK \cite{mccormick2014itk}.
No negative values of $\det(J)$ are observed for the proposed method as can be seen in Table~\ref{table:leftacdcscd}.

We also evaluated the performance of the proposed algorithm in terms of the reliability function $R(d)$ \cite{ayed2009left}. The corresponding $R(d)$ are plot as a function of Dice score in Fig. \ref{fig:reliabilirty}. Our algorithm led to a higher reliability curve on all ACDC, SCD, and left atrium datasets. 

\begin{figure*}
\begin{minipage}[b][1cm]{0.147\linewidth}
		\centering
		\centerline{\includegraphics[trim={2cm 1cm 2cm 1cm},clip, width=\textwidth]{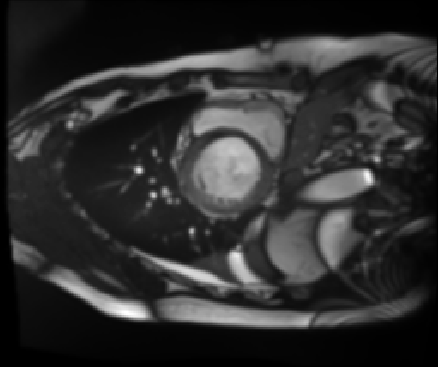}}
	\end{minipage}
	\hfill
	\begin{minipage}[b]{0.147\linewidth}
		\centering
		\centerline{\includegraphics[trim={2cm 1cm 2cm 1cm},clip, width=\textwidth]{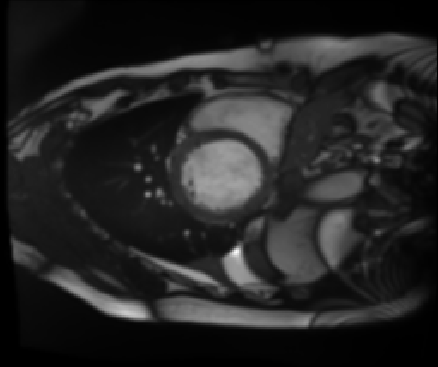}}
	\end{minipage}
	\hfill
	\begin{minipage}[b]{0.147\linewidth}
		\centering
		
		\centerline{\includegraphics[trim={2cm 1cm 2cm 1cm},clip, width=\textwidth]{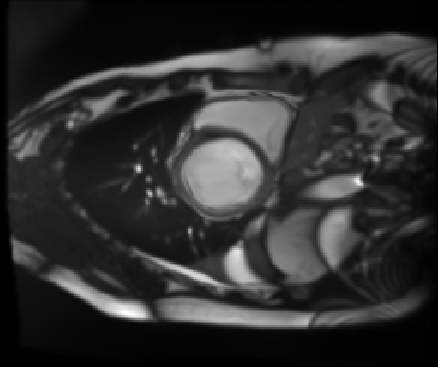}}
	\end{minipage}
	\hfill
	\begin{minipage}[b]{0.147\linewidth}
		\centering
		\centerline{\includegraphics[trim={2cm 1cm 2cm 1cm},clip, width=\textwidth]{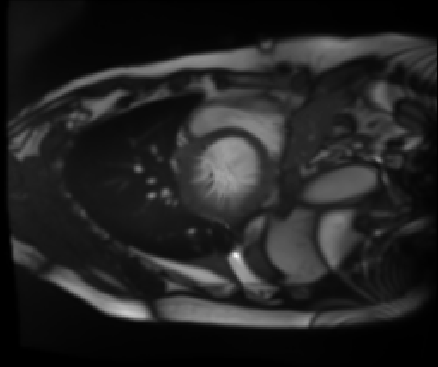}}
	\end{minipage}
	\hfill
	\begin{minipage}[b]{0.147\linewidth}
		\centering
		\centerline{\includegraphics[trim={2cm 1cm 2cm 1cm},clip, width=\textwidth]{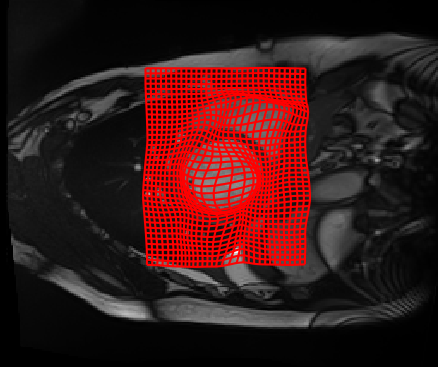}}
	\end{minipage}
	\hfill
	\begin{minipage}[b]{0.147\linewidth}
		\centering
		
		\centerline{\includegraphics[trim={2cm 1cm 2cm 1cm},clip, width=\textwidth]{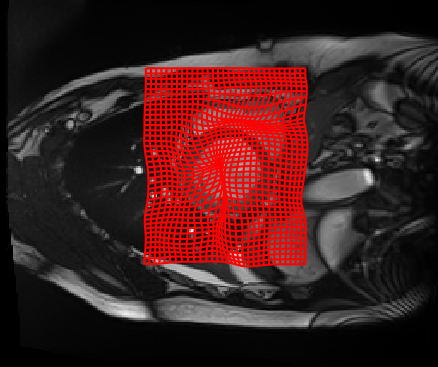}}
	\end{minipage}
	\hfill
	\begin{minipage}[b]{0.99\linewidth}
		\centering
		\centerline{(a) ACDC}\medskip
	\end{minipage}
	
	\begin{minipage}[b]{0.147\linewidth}
	\centering
	\centerline{\includegraphics[trim={1.5cm 1.5cm 1.5cm 1.5cm},clip, width=\textwidth]{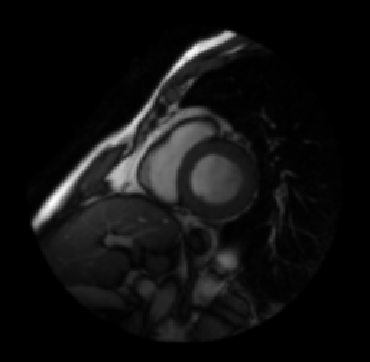}}
	\end{minipage}
	\hfill
	\begin{minipage}[b]{0.147\linewidth}
		\centering
		
		\centerline{\includegraphics[trim={1.5cm 1.5cm 1.5cm 1.5cm},clip, width=\textwidth]{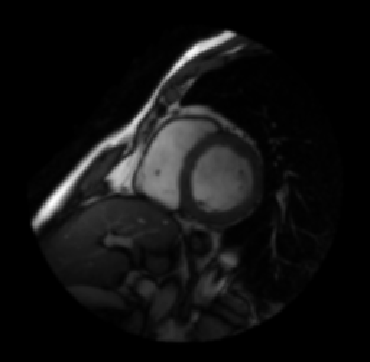}}
	\end{minipage}
	\hfill
	\begin{minipage}[b]{0.147\linewidth}
		\centering
		
		\centerline{\includegraphics[trim={1.5cm 1.5cm 1.5cm 1.5cm},clip, width=\textwidth]{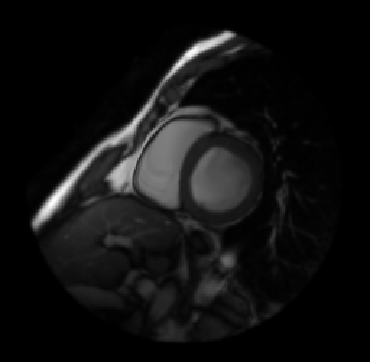}}
	\end{minipage}
	\hfill
	\begin{minipage}[b]{0.147\linewidth}
		\centering

		\centerline{\includegraphics[trim={1.5cm 1.5cm 1.5cm 1.5cm},clip, width=\textwidth]{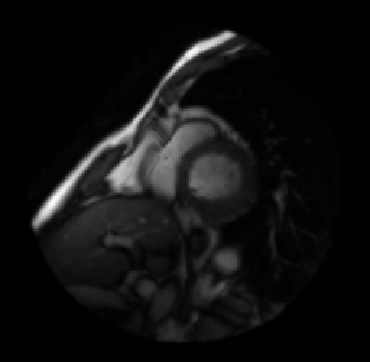}}
	\end{minipage}
	\hfill
	\begin{minipage}[b]{0.147\linewidth}
		\centering
		
		\centerline{\includegraphics[trim={1.5cm 1.5cm 1.5cm 1.5cm},clip, width=\textwidth]{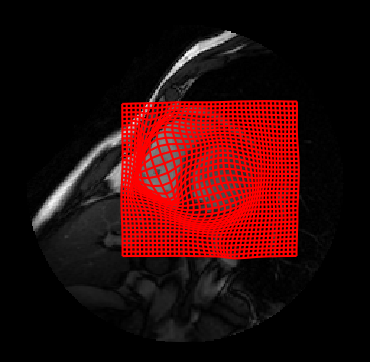}}
	\end{minipage}
	\hfill
	\begin{minipage}[b]{0.147\linewidth}
		\centering
		
		\centerline{\includegraphics[trim={1.5cm 1.5cm 1.5cm 1.5cm},clip, width=\textwidth]{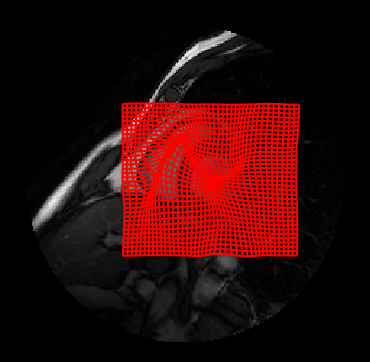}}
	\end{minipage}
	\begin{minipage}[b]{0.99\linewidth}
		\centering

		\centerline{(b) SCD}\medskip
	\end{minipage}
	\begin{minipage}[b]{0.147\linewidth}
	\centering
	
	\centerline{\includegraphics[trim={2cm 0.75cm 2cm 0.75cm},clip, width=\textwidth]{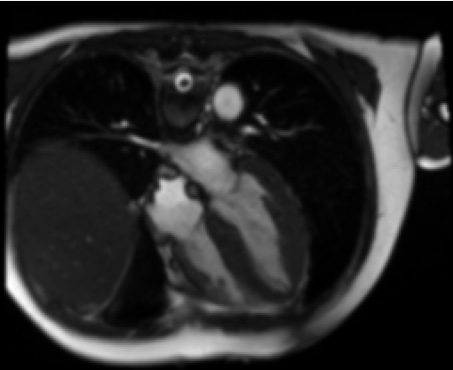}}
	\end{minipage}
	\hfill
	\begin{minipage}[b]{0.147\linewidth}
		\centering
		
		\centerline{\includegraphics[trim={2cm 0.75cm 2cm 0.75cm},clip, width=\textwidth]{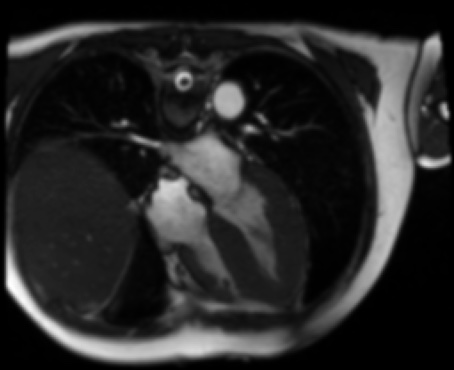}}
	\end{minipage}
	\hfill
	\begin{minipage}[b]{0.147\linewidth}
		\centering
		
		\centerline{\includegraphics[trim={2cm 0.75cm 2cm 0.75cm},clip, width=\textwidth]{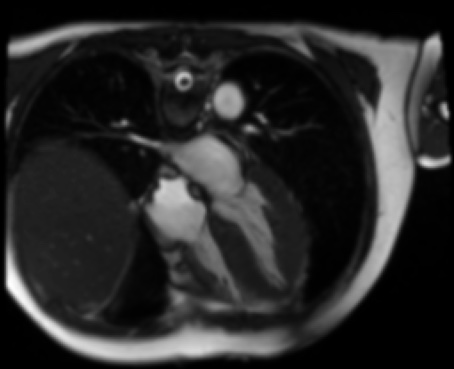}}
	\end{minipage}
	\hfill
	\begin{minipage}[b]{0.147\linewidth}
		\centering
		
		\centerline{\includegraphics[trim={2cm 0.75cm 2cm 0.75cm},clip, width=\textwidth]{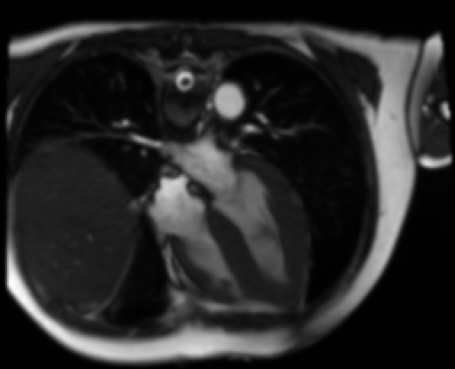}}
	\end{minipage}
	\hfill
	\begin{minipage}[b]{0.147\linewidth}
		\centering
		
		\centerline{\includegraphics[trim={2cm 0.75cm 2cm 0.75cm},clip, width=\textwidth]{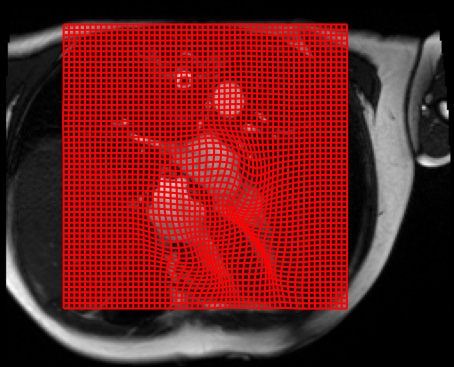}}
	\end{minipage}
	\hfill
	\begin{minipage}[b]{0.147\linewidth}
		\centering
		\centerline{\includegraphics[trim={2cm 0.75cm 2cm 0.75cm},clip, width=\textwidth]{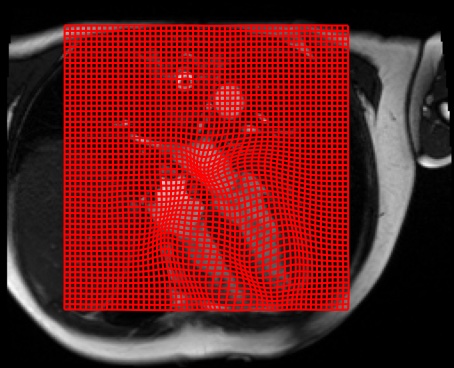}}
	\end{minipage}
	\hspace{0.003mm}
	\begin{minipage}[b]{0.99\linewidth}
		\centering

		\centerline{(c) LA}\medskip
	\end{minipage}
	
	
	\begin{minipage}[b]{0.147\linewidth}
		\centering

		\centerline{ES}\medskip
	\end{minipage}
	\hfill
	\begin{minipage}[b]{0.147\linewidth}
		\centering
	
		\centerline{ED}\medskip 
	\end{minipage}
	\hfill
	\begin{minipage}[b]{0.147\linewidth}
		\centering
	
		\centerline{Registered ES}\medskip
	\end{minipage}
	\hfill
	\begin{minipage}[b]{0.147\linewidth}
		\centering

		\centerline{Registered ED}\medskip
	\end{minipage}
	\hfill
	\begin{minipage}[b]{0.147\linewidth}
		\centering

		\centerline{$D^F$}\medskip
	\end{minipage}
	\hfill
	\begin{minipage}[b]{0.147\linewidth}
		\centering
		\centerline{$D^B$}\medskip
	\end{minipage}
	\caption{Examples of pair images ES and ED from (a) ACDC, (b) SCD and (c) LA. Each row shows forward and backward registered images and the corresponding grid deformations. The registrations are applied to the whole image, however, the grids are cropped to have a better resolution}
	\label{fig:picresults}
	\vspace{-0.2cm}
\end{figure*}

We examined different loss functions, namely, MSE, SSIM and SSIM + MI, in our framework and reported the average time to register one pair of images (Computing both forward and backward deformation fields) as well as the average of DM and HD over $50\%$ of patient datasets in Table~\ref{tale:ACDCscdleftlosses}. It can be seen that the MSE loss function led to the best average time; however, it does not yield the best accuracy in terms of the Dice score. 

We assessed the impact of using different multi-resolution structures on the performance, and reported the corresponding evaluations in terms of DM and HD in Table \ref{table:resoulotions}. The results indicate that the high performance is obtained with the use of two resolutions.

Examples of registered images from ACDC, SCD, and left atrium datasets with their corresponding grid deformation are shown in Fig.~\ref{fig:picresults} (a), (b), and (c), respectively.
Fig.~\ref{fig:picresults} shows the original ED, ES and their correspondence registered images and grid forward and backward deformations over images. The algorithm is applied on whole images; however, to display the deformation tracking on the part that has the most changes, we cropped the grid.

We have also evaluated the outputs from the forward deformation field by ignoring the bidirectional registration to present its effects on the Dice score and HD. The forward deformation field yields Dice score values of $0.88\pm0.17$ and $0.87\pm 0.14$ and HD values of $6.08$ and $5.19$ for ACDC and SCD, respectively, while the proposed bidirectional approach yields Dice score values of $0.92\pm 0.10$ and $0.90\pm 0.15$ and HD values of $4.76$ and $5.06$ for ACDC and SCD, as reported in Table~\ref{table:leftacdcscd}. This clearly indicates the importance of including the bidirectional registration to obtain high performance. 

\section{Conclusion}
We present an ODE-based diffeomorphic recursive framework for multi-resolution deformable registration using a FCN, to estimate the change velocity of forward and backward deformation fields. We have then utilized inverse consistency loss to further guarantee the desirable diffeomorphic properties of the resulting solutions. The proposed framework is learning-free and does not require a dedicated training set.

The proposed approach simultaneously estimates the forward and backward mapping at all levels of the multi-resolution pyramid. 
We evaluated and compared the method using three different MRI datasets against several state-of-the-art traditional and learning-based registration methods. The results demonstrate that our method outperforms both traditional and learning-based methods.
%
%
%

\bibliographystyle{splncs04}
\bibliography{ref}
\end{document}